\documentclass[a4paper,aps,amsfonts,amsmath,nofootinbib,showpacs,byrevtex,showkeys,prd,notitlepage]{revtex4-1}

\usepackage[UKenglish]{babel}
\usepackage{tuemath}

\begin{document}

\title{Dyson--Schwinger approach to Hamiltonian Quantum Chromodynamics}

\author{Davide R.~Campagnari}
\author{Hugo Reinhardt}
\affiliation{Institut f\"ur Theoretische Physik, Eberhard-Karls-Universit\"at T\"ubingen,
Auf der Morgenstelle 14, 72076 T\"ubingen, Germany}

\pacs{11.10.Ef, 12.38.Aw, 12.38.Lg}
\keywords{Hamiltonian approach, variational principle, Coulomb gauge}

\begin{abstract}
The general method for treating non-Gaussian wave functionals in the Hamiltonian
formulation of a quantum field theory, which was previously proposed and developed for
Yang--Mills theory in Coulomb gauge, is generalized to full QCD. For this purpose the quark part
of the QCD vacuum wave functional is expressed in the basis of coherent fermion states,
which are defined in term of Grassmann variables. Our variational ansatz for the QCD
vacuum wave functional is assumed to be given by exponentials of polynomials in the
occurring fields and, furthermore, contains an explicit coupling of the quarks to the gluons. Exploiting Dyson--Schwinger
equation techniques, we express the various $n$-point functions, which are required for the expectation
values of observables like the Hamiltonian, in terms of the variational kernels of our trial ansatz.
Finally the equations of motion for these variational kernels are derived by minimizing the
energy density.
\end{abstract}

\maketitle


\section{Introduction}

One of the major challenges of theoretical particle physics is the understanding
of the low-energy sector of Quantum Chromodynamics (QCD). Despite many years of intensive research,
a thorough and unified picture of the low-energy phenomena of strong interactions,
i.e.,~confinement and spontaneous breaking of chiral symmetry, is still
lacking. Much insight has been gained by 
means of lattice Monte-Carlo calculations, in particular on the gluon sector of QCD.
However, despite much progress, the treatment of dynamical chiral quarks is still a
challenge for the lattice approach, which furthermore struggles to describe the phase diagram of QCD at finite
baryon density due to the notorious sign problem. In addition and in general, a thorough 
understanding of physical phenomena cannot be achieved by numerical lattice simulations 
alone but analytic methods, albeit approximate ones, are needed as well.

Over the last decade substantial efforts have been undertaken to develop non-perturbative
continuum approaches to QCD. These approaches are based on functional methods and can be 
roughly divided into three classes:
i) Dyson--Schwinger equations (DSEs) in Landau \cite{Alkofer:2000wg,*Fischer:2006ub,Fischer:2008uz,Binosi:2009qm}
and Coulomb gauges \cite{Zwanziger:1998ez,Watson:2006yq,Watson:2007mz,*Watson:2007vc,*Popovici:2008ty,Reinhardt:2008pr,Watson:2008fb,*Watson:2010cn,*Watson:2011kv,*Watson:2012ht,Popovici:2010mb,*Popovici:2010ph,*Popovici:2011yz}, 
ii) functional renormalization group (FRG) flow equations \cite{Pawlowski:2005xe,Gies:2006wv,Braun:2011pp} and 
iii) the variational approach to Hamiltonian QCD \cite{Schutte:1985sd,*Szczepaniak:2001rg,Feuchter:2004mk,*Reinhardt:2004mm}. 
These three approaches are intimately related. The first two approaches are based on the functional
integral formulation of QCD in either Landau or Coulomb gauge, while the variational approach
has been mainly applied to the Hamiltonian formulation in Coulomb gauge but has been recently
also extended to the effective action of the functional integral formulation in Landau gauge \cite{Quandt:2013wna,*Quandt:2015aaa}. The equations
of motion of the first two approaches are in fact very similar and in a certain approximation 
(replacing the renormalization group scale $k$ in the loop integrals by its infrared value $k = 0$)
the FRG flow equation becomes a Dyson--Schwinger equation. The FRG flow equations have been
also applied to the Hamiltonian approach in Coulomb gauge and results similar to those 
in the variational treatment were found \cite{Leder:2010ji,Leder:2011yc}. Furthermore, the DSE techniques can be very advantageously
exploited to carry out the variational approach with non-Gaussian wave functionals describing 
interacting quantum fields \cite{Campagnari:2010wc}. In the present paper we use the Dyson--Schwinger equations
to develop a variational approach to the Hamiltonian formulation of QCD.

Previous variational studies within the Hamiltonian approach have focused on the Yang--Mills sector 
and used Gaussian-type ans\"atze for the vacuum wave functional. This has provided a decent
description of both the infrared (IR) and ultraviolet (UV) sector in rough agreement with
the existing lattice data. In particular, a linearly rising static quark potential \cite{Epple:2006hv,Pak:2009em} and a perimeter
law \cite{Reinhardt:2007wh} for the 't Hooft loop \cite{tHooft:1977hy} were found, which are both features of the confined phase. Also the 
connection to the dual Meissner effect (an appealing picture of confinement) was established \cite{Reinhardt:2008ek}.
More recently, the deconfinement phase transition was studied at finite temperatures \cite{Reinhardt:2011hq,Heffner:2012sx}
and the effective potential of the Polyakov loop was calculated \cite{Reinhardt:2012qe,Reinhardt:2013iia}. The obtained critical
temperature is in reasonable agreement with the lattice data. Furthermore, the order of the phase
transition was correctly reproduced \cite{Reinhardt:2013iia}. In the zero-temperature calculation the obtained static gluon 
propagator agrees with the lattice data in the IR and UV but misses some strength in the 
mid-momentum regime. This missing strength can be attributed to the absence of non-Gaussian
terms in the vacuum wave functional \cite{Campagnari:2010wc}.

Generally, Gaussian wave functionals describe quantum field theories in the independent
(quasi-)particle approximation, while truly interacting quantum fields possess
non-Gaussian vacuum wave functionals. In Ref.~\cite{Campagnari:2010wc} a general method
for treating non-Gaussian trial wave functionals in a quantum field theory was proposed.
This method relies on Dyson--Schwinger-type of equations to express the various $n$-point
functions of the quantum field in terms of the variational kernels contained in the
exponent of the ansatz for the vacuum wave functional. So far this method has been
formulated for Bose fields only and was applied to the Yang--Mills sector of QCD using a 
wave functional which includes, besides the usual quadratic term of the Gaussian, also cubic and
quartic terms of the gauge field. In principle, the approach put forward in Ref.~\cite{Campagnari:2010wc}
is general enough to deal with any interacting quantum field theory. In the present paper we 
extend this approach to full QCD. The central point will be the treatment of fermion fields 
interacting with Bose (gauge) fields. To exploit the Dyson--Schwinger equation techniques, the second
quantization of the fermion sector of the theory has to be formulated in terms of Grassmann 
variables. For this purpose we express the quark part of the QCD vacuum wave functional
in terms of coherent fermion states.
We will formulate the present approach to Hamiltonian QCD for general wave functionals but
work out the Dyson--Schwinger-type of equations only for those wave functionals whose
exponent is bilinear in the quark field.

The QCD vacuum wave functional is chosen as the exponential of some polynomial functional of the quark
and gluon fields. The coefficient functions of the various polynomial terms are
treated as variational kernels. By means of Dyson--Schwinger equation techniques we
express the various $n$-point functions, needed for the vacuum expectation value of
observables like the Hamiltonian, in terms of these variational kernels, which in this
context figure as bare vertices. The resulting
equations are different from the usual DSEs, which relate the various \emph{full} (\emph{dressed})
propagators and vertices to the \emph{bare} (inverse) propagators and vertices
occurring in the classical action, and are termed canonical recursive DSEs (CRDSEs)
in the following. By means of the CRDSEs we express the vacuum expectation value
of the QCD-Hamiltonian as functional of the variational kernels of our vacuum wave functional.
Minimization of the energy density with respect to these variational kernels results
then in a set of equations of motion (referred to as `gap equations'), which have to be solved together
with the CRDSEs.

The organization of the rest of the paper is as follows: In Sec.~\ref{sec:coh} we briefly summarize the basic
ingredients of the formulation of the second quantization in terms of Grassmann variables
and present the quark part of the QCD wave functional in the basis of coherent fermion states.
In Sec.~\ref{sec:hamdses} we derive the general form
of the CRDSEs for the static Green functions of QCD, assuming a QCD vacuum wave functional
which in particular contains the coupling between quarks and gluons.
In Sec.~\ref{sec:qcdham}, after introducing the QCD Hamilton operator in Coulomb gauge,
we calculate the energy density in the vacuum state. The variational principle is carried out
in Sect.~\ref{sec:var_prin}, where we derive the equations of motion for the variational kernels of our QCD vacuum wave functional.
A short summary and our conclusions are given in Sect.~\ref{sec:conc}. Here we also briefly
discuss further applications of the approach developed in this work.


\section{\label{sec:coh}Coherent State Description of the Fermionic Fock Space}

As demonstrated in Ref.~\cite{Campagnari:2010wc} for Yang--Mills theory, the use of
non-Gaussian wave functionals in the Hamiltonian approach can be conveniently
accomplished by exploiting Dyson--Schwinger equation techniques known from the Lagrangian
(functional integral) formulation of quantum field theory. This refers, in particular,
to wave functionals describing interacting fields. To
exploit the DSE techniques in the Hamiltonian formulation of QCD it is necessary to
represent the quark operators and wave functionals in terms of anti-commuting Grassmann
fields. In this representation the matrix
elements between Fock-space states are then given by functional integrals over Grassmann
fields. The formulation of the second quantization in terms of Grassmann variables
becomes particularly efficient when coherent fermion states are used.
Below we will briefly review the basic ingredients of the coherent fermion state
representation of Fock space and apply it to Dirac fermions. 


\subsection{\label{sec:rep}Coherent Fermion States and Grassmann Variables}

Consider a Fermi system described in second quantization in terms of creation and annihilation
operators $b^\dagger_k$, $b_k$, satisfying the usual anti-commutation relations
\[
\anticomm{b_k}{b_l} = 0 = \anticomm{b^\dagger_k}{b^\dagger_l} , \qquad
\anticomm{b_k}{b^\dagger_l} = \delta_{kl} ,
\]
where the subscripts $k$, $l$, \ldots\ denote a complete set of single-particle states. 
Let $\ket{0}$ be the Fock vacuum, i.e.
\[
b_k \ket{0} = 0 .
\]
The coherent fermion states $\ket{\zeta}$ are defined as eigenstates of the annihilation
operators \cite{Reinhardt:qm2}
\be\label{159-3}
b_k \ket{\zeta} = \zeta_k \ket{\zeta} ,
\ee
where the $\zeta_k$ are anti-commuting (Grassmann) variables $\anticomm{\zeta_k}{\zeta_l} = 0$.
The corresponding bra-vectors are defined as left eigenstates of the creation operators
\be\label{171-5}
\bra{\zeta} b^\dagger_k = \bra{\zeta} \zeta^*_k \, ,
\ee
where the operation `$*$' denotes the involution. We will keep the same symbol `$*$'
also for the usual complex conjugation of ordinary complex numbers.

The coherent fermion states defined by Eqs.~\eqref{159-3} and \eqref{171-5}
can be expressed in Fock space as 
\be\label{176-7}
\ket{\zeta} = \exp (\zeta \cdot b^\dagger) \ket{0}, \quad \quad
\bra{\zeta}   =  \bra{0} \exp (b \cdot \zeta^*) .
\ee
Here we have skipped the indices and used the short-hand notation 
\[
\eta^* \cdot \zeta = \sum_k \eta^*_k \zeta_k .
\]
From the representation \Eqref{176-7} one easily finds for the scalar product of two coherent states 
\[
\braket{\eta}{\zeta} = \e^{\eta^* \cdot \zeta} .
\]
The coherent fermion states form an over-complete basis of the Fock space, in which the
unit operator has the representation
\be
\label{187-9}
\id = \int \d\zeta^* \d\zeta \: \e^{- \zeta^* \cdot \zeta} \ket{\zeta} \bra{\zeta} ,
\qquad \text{with } \d\zeta^* \d\zeta \equiv \prod_k \d\zeta^*_k \d\zeta_k \,  .
\ee

An arbitrary state $\ket{\varPhi}$ of the Fock space can be expressed in the basis of
coherent fermion states $\ket{\zeta}$ by taking the scalar product
\be\label{rep6}
\varPhi(\zeta^*) = \braket{\zeta}{\varPhi} ,
\ee
which can be interpreted as the ``coordinate representation'' of fermion states with
the Grassmann variables interpreted as classical fermion coordinates. From the
representation \Eqref{176-7} it is clear that the $\varPhi(\zeta^*)$ are functionals of
the $\zeta^*_k$, which can be Taylor expanded 
\be\label{219-14}
\varPhi(\zeta^*) = \sum_n \sum_{k_1 \dots k_n} \varPhi_{k_1 \dots k_n} \zeta^*_{k_1} \dots \zeta^*_{k_n} \, ,
\ee
where the $\varPhi_{k_1 \dots k_n}$ are complex numbers. The representation of
bra vectors is obtained by taking the adjoint of \Eqref{rep6}\footnote{%
The adjoint means the involution for the Grassmann variables and complex conjugation
for ordinary complex numbers.}
\be\label{XX1}
\braket{\varPhi}{\zeta}= \bigl(\varPhi (\zeta^*)\bigr)^* \equiv \varPhi^{*}(\zeta)
\ee
and is obviously a function of the variables $\zeta_k$. From \Eqref{219-14} we find
\[
\varPhi^*(\zeta) = \sum_n \sum_{k_1 \dots k_n} \varPhi^*_{k_1 \dots k_n} \zeta_{k_n} \dots \zeta_{k_1} .
\]
The Fock-space states are given by functions of the creation operators
$f (b^\dagger)$ acting on the vacuum state
\[
\ket{\varPhi} = f (b^\dagger) \ket{0} .
\]
From \Eqref{171-5} we find then the corresponding coherent state representation 
\[
\varPhi (\zeta^*) = f(\zeta^*) ,
\]
where we have used $\braket{\zeta}{0} = 1$, which follows immediately from \Eqref{176-7}.
The scalar product between two Fock states $|\varPhi \rangle$ and $|\varPsi\rangle$ is
easily obtained in the basis of coherent states by inserting the completeness relation \Eqref{187-9}
\be\label{294G3}
\braket{\varPsi}{\varPhi} = \int \d\zeta^* \d\zeta \: \e^{- \zeta^* \cdot \zeta} \, \varPsi^{*}(\zeta) \, \varPhi(\zeta^*) ,
\ee
where we have also used the definitions \eqref{rep6} and \eqref{XX1}.
Using \Eqref{171-5} and
\[
\bra{\zeta} b_k = \frac{\partial}{\partial \zeta^*_k} \: \bra{\zeta} 
\]
we find for the action of an operator on a Fock state in the coherent state basis
\[
\bra{\zeta} O (b, b^\dagger) \ket{\varPhi} = O \bigl(\tfrac{\partial}{\partial \zeta^*}, \zeta^* \bigr) \, \varPhi (\zeta^*) 
\]
and similarly for matrix elements between states of Fock space
\be\label{268-23}
\bra{\varPsi} O (b, b^\dagger) \ket{\varPhi} = \int \d\zeta^* \d\zeta \: \e^{- \zeta^* \cdot \zeta} \,
\varPsi^*(\zeta) \, O \bigl(\tfrac{\partial}{\partial \zeta^*}, \zeta^* \bigr) \, \varPhi (\zeta^*) .
\ee
In the next subsection the coherent fermion state representation of the Fermionic Fock
space given above is extended to Dirac Fermions.


\subsection{\label{sec:cohdir}Coherent-State Representation of Dirac Fermions}

The description of Fermi systems in terms of Grassmann variables outlined above can be
immediately applied to Dirac fermions once the fermion field is expressed in terms of
creation and annihilation operators.

In this paper we use a compact notation in which a single digit 1, 2, \ldots{}
represents all indices of the time-independent field. For example, for the quark field we have
\[
\psi(1) \equiv \psi_{s_1}^{k_1}(\vx_1) ,
\]
where $s_1$ denotes the Dirac spinor index, while $k_1$ stands for the colour (and
possibly flavour) index.\footnote{In the present paper the quark flavour will be irrelevant,
but the subsequent considerations do not change when flavour is included.}
A repeated index implies integration of spatial coordinates and summation over the discrete
indices (spinor, colour, flavour). Furthermore, we define the Kronecker symbol in the numerical
indices to contain besides the usual Kronecker symbol for the discrete indices also the
$\delta$ function for the continuous coordinates, e.g.
\[
\delta(1,2) = \delta(\vx_1-\vx_2) \, \delta_{s_1s_2} \delta^{k_1k_2} \dots
\]
The usual anticommutation relation for the Dirac field reads
\be\label{X0}
\anticomm{\psi(1)}{\psi^\dag(2)} = \delta(1,2) .
\ee
Let
\be\label{h0}
h_0(1,2) = \delta^{m_1m_2} \bigl( -\I*\vec{\alpha}\cdot\grad + \beta m \bigr) \delta(\vx_1-\vx_2)
\ee
denote the Dirac Hamiltonian of free quarks with a bare mass $m$.
This Hamiltonian possesses (an equal number of) positive and negative energy eigenstates.
We can expand the quark field $\psi(1)$ in terms of these eigenstates. Let $\psi_{\pm}(1)$
denote the part formed from the positive/negative energy modes. Obviously, we have
\[
\psi(1) = \psi_+(1) + \psi_-(1) .
\]
For the subsequent considerations it will be convenient to introduce orthogonal
projectors $\Lambda_\pm$ onto the positive and negative energy part of the Dirac field,
i.e.
\be\label{1447X1}
\psi_\pm(1) = \Lambda_\pm(1,2) \psi(2),
\ee
satisfying
\be\label{1447X1B}
\Lambda_+(1,2) + \Lambda_-(1,2) = \delta(1,2), \qquad
\Lambda_+(1,2)\Lambda_-(2,3) = 0, \qquad
\Lambda_\pm(1,2) \, \Lambda_\pm(2,3) = \Lambda_\pm(1,3).
\ee
These projectors can be expressed by spectral sums over the positive and negative,
respectively, energy modes of the Hermitian Dirac Hamiltonian $h_0(1,2)$, and satisfy
$\Lambda_\pm^\dag(1,2) = \Lambda_\pm(1,2)$.
With this relation we find for the adjoint fermion operator from \Eqref{1447X1}
\be\label{1447X1A}
\psi_\pm^\dag(1) = \psi^\dag(2) \Lambda_\pm(2,1) .
\ee
From Eqs.~\eqref{1447X1} and \eqref{1447X1A} it follows with \eqref{1447X1B}
\[
\anticomm{\psi_\pm(1)}{\psi_\pm^\dag(2)} = \Lambda_\pm(1,2),
\qquad \anticomm{\psi_\pm(1)}{\psi_\mp^\dag(2)} = 0.
\]
In the bare (free Dirac) vacuum state $\ket{0}$ all negative energy modes are filled,
while the positive energy modes are empty, implying
\be\label{mq}
\psi_+(1) \ket{0} = 0 = \psi_-^\dag(1) \ket{0} .
\ee
In analogy to \Eqref{159-3} we define coherent fermion states $\ket{\xi}\equiv\ket{\xi_+,\xi_-}$
by
\be\label{1447X3}
\psi_+(1) \ket{\xi} = \xi_+(1) \ket{\xi} , \qquad
\psi_-^\dag(1) \ket{\xi} = \xi_-^\dag(1) \ket{\xi} ,
\ee
where $\xi_\pm$ are Grassmann fields. Since $\psi_\pm(1) = \Lambda_\pm(1,2) \psi_\pm(2)$,
the Grassmann fields also satisfy
\[
\xi_\pm(1) = \Lambda_\pm(1,2) \xi_\pm(2)
\]
and the positive- and negative-energy component fields can be assembled into a single
Grassmann-valued spinor field
\be\label{49X5A}
\xi(1) = \xi_+(1) + \xi_-(1), \qquad \xi_\pm(1) = \Lambda_\pm(1,2) \xi(2)
\ee
satisfying
\be\label{1447G4}
\frac{\delta \xi_\pm(1)}{\delta \xi_\pm(2)} = \Lambda_\pm(1,2) , \qquad
\frac{\delta \xi^\dag_\pm(1)}{\delta \xi^\dag_\pm(2)} = \Lambda_\pm(2,1) .
\ee
The coherent fermion states defined by \Eqref{1447X3} have the Fock-space representation
[cf.~\Eqref{176-7}]
\[
\ket{\xi} = \exp\left[ \xi_+(1) \psi^\dag_+(1) + \xi_-^\dag(1) \psi_-(1) \right] \ket{0}
\]
from which follows
\be\label{49X5}
\psi^\dag_+(1) \ket{\xi} = \frac{\delta}{\delta\xi_+(1)} \ket\xi , \qquad
\psi_-(1) \ket{\xi} = \frac{\delta}{\delta\xi_-^\dag(1)} \ket\xi .
\ee

Using Eqs.~\eqref{1447X3} and \eqref{49X5} 
the action of Dirac field operators on fermion states $\lvert\varPhi\rangle$
is expressed in the basis of coherent states as
\be\begin{split}\label{vev3}
\bra{\xi} \psi(1) \ket{\varPhi} &= \left( \xi_-(1) + \frac{\delta}{\delta \xi_+^\dag(1)}\right) \varPhi[\xi_+^\dag,\xi_-] , \\
\bra{\xi} \psi^\dag(1)  \ket{\varPhi} &= \left( \xi_+^\dag(1) + \frac{\delta}{\delta \xi_-(1)} \right) \varPhi[\xi_+^\dag,\xi_-] ,
\end{split}\ee
where $\braket{\xi}{\Phi} \equiv \varPhi[\xi_+^\dag,\xi_-]$ is the coherent-state representation of the
quark vacuum wave functional $\lvert\varPhi\rangle$, which can be interpreted as the ``coordinate representation''
of the latter. Notice that from the definition \Eqref{1447X3} of the coherent fermion states
follows that the quark vacuum wave functional depends only on $\xi_+^\dag$ and $\xi_-$.

In analogy to \Eqref{268-23} the matrix element of an operator $\calO$ between
fermionic Fock states $|\varPhi_1\rangle$ and $|\varPhi_2\rangle$
is given by
\be\label{dir4}
\bra{\varPhi_1} \calO[\psi,\psi^\dag] \ket{\varPhi_2} = \int \calD\xi^\dag \, \calD\xi \:
\e^{-\mu} \: \varPhi_1^*[\xi_+,\xi_-^\dag] \:
\calO \bigl[ \xi_-+\tfrac{\delta}{\delta \xi_+^\dag},\xi_+^\dag + \tfrac{\delta}{\delta \xi_-}\bigr] \:
\varPhi_2[\xi_+^\dag,\xi_-] ,
\ee
where we have introduced the quantity
\be\label{dir4a}
\mu = \xi^\dag_+(1) \, \xi_+(1) - \xi^\dag_-(1) \, \xi_-(1) ,
\ee
arising from the integration measure of the Grassmann fields [cf.~\Eqref{294G3}].
Note that $\mu$, being bilinear in the Grassmann variables,
commutes with any element of the Grassmann algebra. This quantity can be rewritten
with the help of Eqs.~\eqref{1447X1B} and \eqref{49X5A} as
\be\label{mu-def}
\mu = \xi^\dag(1) \bigl[ \Lambda_+(1,2) - \Lambda_-(1,2) \bigr] \xi(2) \equiv 
\xi^\dag(1) \, Q_0^{-1}(1,2) \, \xi(2) ,
\ee
where the quantity
\be\label{bareprop}
Q_0^{-1}(1,2) = \Lambda_+ (1,2) - \Lambda_-(1,2) = Q_0(1,2)
\ee
represents the free static quark propagator
\be\label{barepropG6}
Q_0(1,2) = \bra{0} \comm{\psi(1)}{\psi^\dag(2)} \ket{0} ,
\ee
with $\ket{0}$ being the Fock vacuum \Eqref{mq} of the free quarks, as we will discuss
in more detail at the end of Sec.~\ref{sec:qdse}.


\subsection{Coordinate Representation of QCD Wave Functionals}

With the coherent states $\ket{\xi}$ [\Eqref{1447X3}] of the Dirac fermions at hand
we are now in a position to express QCD wave functionals in the coordinate representation.  The coordinates
of the gauge field are its spatial components $A_i^a(\vx)$.
In analogy to the fermion field we will collect all indices of the gluon field in a single
digit $A(1)\equiv A_{i_1}^{a_1}(\vx_1)$, where $a_1$ denotes the colour index of the adjoint
representation and $i_1$ is a spatial Lorentz index. In the coordinate representation of the 
wave functional of the Yang--Mills sector $\varPsi [A] = \braket{A}{\varPsi}$ we have for 
the operators of the canonical variables
\[
\bra{A} \hat{A}(1) \ket{\varPsi} = A(1) \, \varPsi[A] , \qquad
\bra{A} \hat{\Pi}(1) \ket{\varPsi} = \frac{\delta \varPsi[A]}{\I \, \delta A(1)}.
\]
We will work here in Coulomb gauge $\partial_i A_i^a=0$, where only the transverse
components of the gauge field are left. In our compact notation we have
\be\label{xG6}
\frac{\delta A(1)}{\delta A(2)} = t(1,2) ,
\ee
where
\[
t(1,2) = \delta^{a_1 a_2} \left( \delta_{i_1i_2} - \frac{\partial_{i_1}\partial_{i_2}}{\partial^2} \right) \delta(\vx_1-\vx_2)
\]
is the transverse projector.

The vacuum state of QCD can be written in the coordinate representation (i.e.,~coherent-state representation
for the fermions) as
\be\label{qcd0}
\varPsi[A,\xi^\dag_+,\xi_-] \eqcolon \exp\left\{ -\frac12 \, S_A[A] - S_f[\xi^\dag_+,\xi_-,A] \right\} ,
\ee
where $S_A$ defines the vacuum wave functional of pure Yang--Mills theory, while $S_f$ defines
the wave functional of the fermions interacting with the gluons. On general grounds
$S_f$ contains only even powers of Grassmann variables
so that this quantity, as well as the vacuum wave functional $\varPsi[A,\xi^\dag_+,\xi_-]$,
commutes with any Grassmann field.

A comment is here in order concerning the representation \Eqref{qcd0} of the vacuum wave
functional for the gluon and quark fields. For the bosonic gluon field we use here the
usual ``coordinate'' representation regarding its spatial components as the (classical)
coordinates of the theory. As already discussed in Sec.~\ref{sec:rep},
the classical analogues of the fermion fields are the Grassmann
variables and the ``coordinate representation'' of the fermion wave functional is the
coherent (-fermion) state representation, \Eqref{rep6}.
We could also use bosonic coherent states for the gluons but this is not necessary and
we will not use it since the usual coordinate representation is in this case quite convenient.

For sake of illustration we quote the expressions for the perturbative QCD vacuum state
\cite{Campagnari:2009km,*Campagnari:2009wj,*Campagnari:2014hda}.
The perturbative Yang--Mills vacuum is obtained by choosing in \Eqref{qcd0} the quadratic
gluonic ``action''
\be\label{G5}
S_A = \int \frac{\mathrm{d}^3 p}{(2\pi)^3} \: A_i^a(\vp) \, \abs{\vp} \, A_i^a(-\vp) .
\ee
On the other hand, the ground state $\ket{0}$ [\Eqref{mq}] of the free (perturbative)
quarks reads in the coherent state basis $\varPhi[\xi_+^\dag,\xi_-]=\braket{\xi}{0}=1$
(i.e.,~$S_f[\xi_+^\dag,\xi_-,A]=0$). This is because the kinematics of the free fermions
is already encoded in the integration measure $\mu$ of the Grassmann fields,
see Eqs.~\eqref{dir4}--\eqref{mu-def}.

Eventually we are interested in the Hamiltonian formulation of QCD in Coulomb gauge. The
coordinate representation of gluonic states and matrix elements has been presented in Ref.~\cite{Campagnari:2010wc}.
Writing the QCD vacuum wave 
functional in the form (\ref{qcd0}) and using the representation (\ref{dir4}) of 
the fermionic matrix elements, expectation values in gauge-fixed QCD are given by\footnote{%
With a slight abuse of notation we will use the same symbol for the gauge field operator
and for the field variable to be integrated over. It should be always clear from the context
which quantity is meant.}
\be\label{qcd1}
\vev*{\calO[\psi,\psi^\dag,A,\Pi]} = \int \calD\xi^\dag \, \calD\xi \, \calD A \: \calJ_A \e^{-\mu}
\,\e^{-\frac12 S_A-S_f^*} \:
\calO\bigl[\xi_-+\tfrac{\delta}{\delta\xi_+^\dag},\, \xi_+^\dag+\tfrac{\delta}{\delta\xi_-},\, A,\, \tfrac{\delta}{\I*\delta A} \bigr] \:
\e^{-\frac12 S_A-S_f} ,
\ee
where
\be\label{fp}
\calJ_A = \Det G_A^{-1}
\ee
is the Faddeev--Popov determinant. In Coulomb gauge the Faddeev--Popov operator reads
\be\label{xyz}
G^{-1}_A = \bigl( -\delta^{ab} \, {\partial^2_{\vx}} - g \, f^{acb} A^c_i(\vx) \, 
\partial_i^{\vx} \bigr) \delta(\vx-\vy) \, .
\ee
Here $g$ is the bare coupling constant and $f^{acb}$ are
the structure constants of the $\mathfrak{su}(N)$ algebra. The functional integration
in \Eqref{qcd1} runs over transverse gauge field configurations and is, in principle, restricted to the
first Gribov region or, more precisely, to the fundamental modular region.

Once the functional derivatives in \Eqref{qcd1} are taken, the vacuum expectation value
of an operator boils down to a functional average of the form
\be\label{dir4c}
\vev{f[A,\xi,\xi^\dag]}
= \int \calD\xi^\dag \, \calD\xi \, \calD A \: \calJ_A \: \e^{-S_A - S_f - S_f^* - \mu} \: f[A,\xi,\xi^\dag],
\ee
where $f$ is a functional of the fields only (i.e.,~$f$ contains no functional derivatives).

For the variational approach to QCD to be developed later we need the vacuum expectation
values of products of gluon operators $A$, $\Pi$, and, in particular, quark field operators
$\psi$, $\psi^\dag$. Furthermore,
to exploit DSEs techniques we have to express these expectation values by
$n$-point functions of the Grassmann fields $\xi$, $\xi^\dag$. This can be achieved by
means of \Eqref{vev3}. To illustrate how this is accomplished we consider temporarily the
quark sector only and omit the
integration over the gauge field. Using Eqs.~\eqref{dir4} and \eqref{qcd0} we obtain
for the quark bilinear
\[
\vev{ \psi(1) \, \psi^\dag(2) } =
\int {\cal{D}} \xi^\dagger {\cal{D}} \xi \: \e^{-\mu-S_f^*}
\biggl( \xi_-(1)+\frac{\delta}{\delta\xi^\dag_+(1)}\biggr) \biggl( \xi_+^\dag(2)+\frac{\delta}{\delta\xi_-(2)}\biggr) \e^{-S_f} .
\]
Since $S_f^*$ is independent of $\xi_+^\dag$ and $\xi_-$ it is convenient
to perform integrations by parts with respect to $\xi^\dag_+(1)$ and $\xi_-(2)$.\footnote{%
Recall that the formula for integration by parts for Grassmann variables reads
\[
\int \d\eta \: f(\eta) \: \frac{\d g(\eta)}{\d\eta} = - \int \d\eta \: \frac{\d f(-\eta)}{\d\eta} \: g(\eta) .
\]
}
The integration by parts with respect to $\xi_+^\dag(1)$ yields
\[
\vev{ \psi(1) \, \psi^\dag(2) }
= \int {\cal{D}} \xi^\dagger {\cal{D}} \xi  \: \e^{-\mu-S_f^*} \:
 \xi(1) \left( \xi_+^\dag(2)+\frac{\delta}{\delta\xi_-(2)}\right) \e^{-S_f} ,
\]
where we have used the first equation of \Eqref{49X5A}.
In the same way an integration by parts with respect to $\xi_-(2)$ yields
\begin{align}
\vev{ \psi(1) \, \psi^\dag(2) } &= \int 
{\cal{D}} \xi^\dagger {\cal{D}} \xi \: 
\e^{-S_f^*-\mu}
\left(\xi(1) \, \xi_+^\dag(2) + \frac{\delta\xi(1)}{\delta\xi_-(2)} + \xi(1) \: \frac{\delta\mu}{\delta\xi_-(2)} \right) \e^{-S_f}\nn\\
\label{vevs3}
&= 
\int {\cal{D}} \xi^\dagger {\cal{D}} \xi \: \e^{-S_f^*-\mu} \: \xi(1) \, \xi^\dag(2) \, \e^{-S_f} + \Lambda_-(1,2) ,
\end{align}
where again Eqs.~\eqref{49X5A} and \eqref{1447G4} and the normalization of the functional
integral were used. Finally, we can generalize \Eqref{vevs3} 
to the vacuum expectation value of full QCD (i.e.,~taking also the functional average over the gluon 
sector) and include also a functional $f[A]$ of the gauge field. One obtains then
\be\label{vev2}
\vev{ \psi(1) \, \psi^\dag(2) \, f[A] } = \vev{\xi(1) \, \xi^\dag(2) \, f[A] } + \Lambda_{-}(1,2) \vev{f[A]} .
\ee
Using the anticommutation relation \Eqref{X0}, from the last relation follows
\be\label{vev3a}
\vev{ \psi^\dag(1) \, \psi(2) \, f[A] } = \vev{\xi^\dag(1) \, \xi(2) \, f[A] } + \Lambda_{+}(2,1) \vev{f[A]} .
\ee
The expectation value of four fermion operators can be derived along the same lines,
resulting in
\be\begin{split}\label{vev4}
\vev{ \psi^\dag(1) \, \psi(2) \, \psi^\dag(3) \, \psi(4) \, f[A]} = {}
& \vev{\xi^\dag(1) \, \xi(2) \, \xi^\dag(3) \, \xi(4) \, f[A]} + \\
&+ \vev{\xi^\dag(1) \, \xi(2) \, f[A]} \, \Lambda_{+}(4,3) + \vev{\xi(2) \, \xi^\dag(3) \, f[A]} \, \Lambda_{+}(4,1) \\
&+ \vev{\xi^\dag(3) \, \xi(4) \, f[A]} \, \Lambda_{+}(2,1) + \vev{\xi^\dag(1) \, \xi(4) \, f[A]} \, \Lambda_{-}(2,3) \\
&+ \bigl[ \Lambda_{+}(2,1) \, \Lambda_{+}(4,3) + \Lambda_{+}(4,1) \, \Lambda_{-}(2,3)\bigr] \vev{f[A]} .
\end{split}\ee


\section{\label{sec:hamdses}Canonical Recursive Dyson--Schwinger Equations of QCD}

As already discussed in the introduction, in the Hamiltonian approach to a quantum field
theory the use of non-Gaussian wave functionals is most conveniently accomplished by
exploiting DSE techniques known from the functional integral approach. In the Hamiltonian
approach, the scalar products or matrix elements between the wave functionals are given
by functional integrals [see \Eqref{qcd1}], which have an analogous structure to those
of vacuum transition amplitudes in the functional integral approach, except for a different
action and that the fields to be integrated over live in the spatial subspace only. In general,
DSEs relate various propagators (or proper $n$-point functions) to each other through the
vertices of the action. In the Hamiltonian approach the ``action'' is defined by the ansatz
for the vacuum wave functional [see \Eqref{qcd0}] and the corresponding generalized
DSEs, the CRDSEs, are used to
express the $n$-point functions, occurring e.g. in the vacuum expectation value of the
Hamiltonian or other observables, in terms of the variational kernels occurring in the
exponent of the wave functional.

\subsection{\label{subsec:qcd}General Form of the CRDSEs}
In the following we derive the CRDSEs for the Hamiltonian
approach to QCD. The general structure of these equations does not depend
on the specific ansatz for the vacuum wave functional. Therefore we will leave the ``action'' 
$\tfrac12 S_A + S_f$ in \Eqref{qcd0} for the moment arbitrary. 

In the preceding section we have seen that any vacuum expectation value of an operator
involving the quark fields $\psi$ and $\psi^\dag$, and the gluon coordinate and momentum operators
$A$ and $\Pi$, can be reduced to an expectation value of a functional of the Grassmann fields
$\xi$, $\xi^\dag$ and the field ``coordinate'' $A$, see \Eqref{dir4c}. For such expectation values we can
write down Dyson--Schwinger-type of equations in the standard fashion by integrating a total derivative. 

Taking derivatives with respect to the fermion fields yields
\begin{subequations}\label{dses5}
\begin{align}
\vev*{ \frac{\delta S}{\delta\xi(1)} \: f[A,\xi,\xi^\dag] } &= \vev*{\frac{\delta f[A,\xi,\xi^\dag]}{\delta\xi(1)}} , \\
\vev*{ \frac{\delta S}{\delta\xi^\dag(1)} \: f[A,\xi,\xi^\dag] } &= \vev*{\frac{\delta f[A,\xi,\xi^\dag]}{\delta\xi^\dag(1)}} .
\end{align}
\end{subequations}
where $f$ is any functional of the fields, and
\be\label{vac7}
S = S_A + S_f + S_f^* + \mu
\ee
is the total ``action'' defined by the vacuum wave functional \Eqref{qcd0} and with $\mu$, defined in \Eqref{dir4a},
arising from the integration measure over the fermion fields.

Similarly, the derivative with respect to the gauge field $A$ leads to
\be\label{dses6}
\vev*{\frac{\delta S}{\delta A(1)} \: f[A,\xi,\xi^\dag] } =
\vev*{\frac{\delta f[A,\xi,\xi^\dag]}{\delta A(1)}} + \widetilde{\Gamma}_0(3,2;1) \, \vev[\Big]{ G_A(2,3) \, f[A,\xi,\xi^\dag] } ,
\ee
where the last term arises from the derivative of the Faddeev--Popov determinant,
Eqs.~\eqref{fp} and  \eqref{xyz}, which gives rise to the bare ghost-gluon vertex
\be\label{bggv}
\widetilde{\Gamma}_0(2,3;1) = \frac{\delta G_A^{-1}(2,3)}{\delta A(1)} \, .
\ee
Equations \eqref{dses5} and \eqref{dses6} are the basic CRDSEs of the Hamiltonian approach
to QCD. By choosing for $f[A,\xi,\xi^\dag]$ succes\-sively higher powers of the fields one
generates from these equations infinite towers of CRDSEs. The three different towers of
equations obtained from Eqs.~\eqref{dses5} and \eqref{dses6} are equivalent to each other
as long as no approximations are introduced.


\subsection{Ansatz for the Vacuum State}

The CRDSEs derived in the previous subsection are quite general, and their structure
does not depend on the details of the specific ansatz for the vacuum wave functional \Eqref{qcd0}.
In order to proceed further we have to specify the form of the vacuum wave functional. 
Since we are mainly interested here in the quark sector, we will use a Gaussian wave
functional for the Yang--Mills sector
\be\label{ans11}
S_A = \omega(1,2) \, A(1) A(2);
\ee
the generalization to non-Gaussian functionals is straightforwardly accomplished by using
the CRDSE approach developed in Ref.~\cite{Campagnari:2010wc} for the Yang-Mills sector.

The fermionic vacuum state is chosen in the form
\be\label{dir5}
S_f = \xi_+^\dag(1) \, K_A(1,2) \, \xi_-(2) = \xi^\dag(1) \, \Lambda_+(1,1') \, K_A(1',2') \, \Lambda_-(2',2) \, \xi(2)
\ee
where the kernel $K_A$ is supposed to contain also the gauge field, so that $S_f$ includes
also the interaction of the quarks with spatial gluons. We choose this kernel in the form
\be\label{ans2}
K_A(1,2) = K_0(1,2) + K(1,2;3) \, A(3) ,
\ee
where $K_0$ and $K$ are the variational kernels with respect to which we will later
minimize the energy. The ansatz \Eqref{ans2} can be considered as the leading order expansion of $K_A$ 
in powers of the gauge field. Furthermore this ansatz specifies the quark wave functional as 
Slater determinant, which guarantees the validity of Wick's theorem in the quark sector.
For later use we also quote the complex  conjugate fermionic functional
\[
S_f^* = \xi^\dag(1) \, \Lambda_-(1,1') \, K_A^\dag(1',2') \, \Lambda_+(2',2) \, \xi(2) \, ,
\]
where
\[
K_A^\dag(1,2) = [K_A(2,1)]^* = K_0^*(2,1) + K^*(2,1;3) \, A(3).
\]
With the ansatz specified by Eqs.~\eqref{dir5} and \eqref{ans2} the fermionic part of the
action \Eqref{vac7}, $S_f+S_f^*+\mu$, can be rewritten as
\be\label{ans7}
S_f+S_f^*+\mu = \xi^\dag(1) \bigl[ Q_0^{-1}(1,2) + \bar\gamma(1,2) + \bar\Gamma_0(1,2;3) \, A(3) \bigr] \xi(2) ,
\ee
where $Q_0$ is defined by \Eqref{bareprop} and we have introduced the biquark kernel
\be\label{xG9}
\bar\gamma(1,2) = \Lambda_{+}(1,1') \, K_0(1',2') \, \Lambda_{-}(2',2) + \Lambda_{-}(1,1') \, K_0^\dag(1',2') \, \Lambda_{+}(2',2)
\ee
and the bare quark-gluon vertex
\be\label{ans9}
\bar\Gamma_0(1,2;3) = \Lambda_{+}(1,1') \, K(1',2';3) \, \Lambda_{-}(2',2) + \Lambda_{-}(1,1') \, 
K^\dag(1',2';3) \, \Lambda_{+}(2',2) .
\ee

With the explicit form of the vacuum wave functional given by Eqs.~\eqref{ans11} and
\eqref{ans7} the CRDSEs \eqref{dses5} and \eqref{dses6} take the following form
\begin{subequations}
\begin{gather}
\label{ans15a}
\bigl[ Q_0^{-1}(1,2) + \bar{\gamma}(1,2) \bigl] \vev{\xi(2) \, f} + \bar\Gamma_0(1,2;3) \vev{\xi(2) \, A(3) \, f} = \vev*{\frac{\delta f}{\delta \xi^\dag(1)}} ,\\
\label{ans15c}
\bigl[ Q_0^{-1}(2,1) + \bar{\gamma}(2,1) \bigl] \vev{\xi^\dag(2) \, f} + \bar\Gamma_0(2,1;3) \vev{\xi^\dag(2) \, A(3) \, f} = -\vev*{\frac{\delta f}{\delta \xi(1)}} ,\\
\label{ans15b}
2 \omega(1,2) \vev{A(2) \, f} - \bar\Gamma_0(3,2;1) \vev{\xi(2) \, \xi^\dag(3) \, f} - \widetilde\Gamma_0(1;3,2) \vev{G_A(2,3) \, f} = \vev*{\frac{\delta f}{\delta A(1)}} .
\end{gather}
\end{subequations}
Choosing the functional $f$ appropriately, these equations allow us to express the
various $n$-point functions of the fields $A$, $\xi$, $\xi^\dag$ in terms of the
(variational) kernels $\omega$ [\Eqref{ans11}], $K_0$ and $K$ [Eqs.~\eqref{dir5} and \eqref{ans2}]
of the vacuum wave functional.
Later on we will use these equations to express the expectation value of the QCD Hamiltonian
in terms of these variational kernels.


\subsection{\label{sec:qdse}Quark Propagator CRDSE}

\begin{figure}
\centering
\includegraphics[height=8ex]{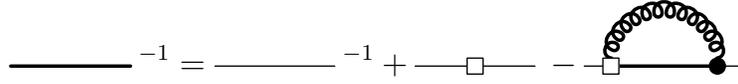}
\caption{\label{fig:fermionDSE}Diagrammatic representation of the CRDSE~\eqref{qdse2} for
the quark propagator. Continuous and wavy lines represent, respectively, quark and gluon
propagators, where thin and thick lines stand, respectively, for bare and fully dressed
propagators. Empty boxes represent variational kernels while full dots stand for
one-particle irreducible vertex functions. A dictionary of our diagrammatic conventions
is given in the appendix.}
\end{figure}

Since our quark wave functional is the exponent of a quadratic form in the fermion fields
[see \Eqref{dir5}] we can express all fermionic vacuum expectation values of these fields
in terms of the fermionic two-point function, which is a manifestation of Wick's theorem.
The CRDSE for the full quark propagator
\be\label{grassprop}
Q(1,2) \coloneq \vev{\xi(1) \, \xi^\dag(2)}
\ee
can be obtained by putting $f=\xi^\dag$ in \Eqref{ans15a}. This yields
\be\label{qdse1}
\bigl[ Q_0^{-1}(1,3) + \bar\gamma(1,3) \bigr] Q(3,2) + \bar{\Gamma}_0(1,3;4) \vev{\xi(3) \xi^\dag(2) A(4)} = \delta(1,2) .
\ee
To resolve the occurring three-point function we introduce the full quark-gluon vertex\footnote{%
We denote quark-gluon vertex functions by $\bar\Gamma$, and ghost-gluon vertex functions
by $\widetilde\Gamma$, see \Eqref{ggv} below.}
$\bar\Gamma$ by
\be\label{qgv}
\vev{A(1) \xi(2) \xi^\dag(3)} \eqcolon -  D(1,1') \, Q(2,2') \, Q(3',3) \bar{\Gamma}(2',3';1') ,
\ee
where
\be\label{gluonprop}
\vev{A(1) \, A(2)} \eqcolon  D(1,2)
\ee
is the static (equal-time) gluon propagator.
With the definitions of the Grassmann propagator [\Eqref{grassprop}] and the
quark-gluon vertex [\Eqref{qgv}] we can cast the CRDSE for the quark propagator
\Eqref{qdse1} into the form
\be\label{qdse2}
Q^{-1}(1,2) = Q_0^{-1}(1,2) + \bar\gamma(1,2) - \bar\Gamma_0(1,3;4) Q(3,3') D(4,4') \bar\Gamma(3',2;4') 
\ee
with $Q_0^{-1}$ defined by \Eqref{bareprop}. Equation \eqref{qdse2} is diagrammatically represented in Fig.~\ref{fig:fermionDSE}.
Alternatively we could have put $f=\xi$ in \Eqref{ans15c}; this would result in the same
CRDSE for the quark propagator as \Eqref{qdse2} except that bare and full quark-gluon vertex
would be interchanged.

Note that the quark two-point function $Q$ defined in \Eqref{grassprop} is not the true
equal-time quark propagator, given in the Hamiltonian approach by
\be\label{qpt4}
S(1,2) \coloneq \frac12 \vev*{\comm{\psi(1)}{\psi^\dag(2)}} .
\ee
Here the commutator arises from the equal-time limit of the time ordering in the
time-dependent Green function. Using Eqs.~(\ref{vev2}) and (\ref{vev3a}) with $f[A]=1$
we find from \Eqref{qpt4} the relation
\be\label{qprop}
S(1,2) = \vev{\xi(1) \, \xi^\dag(2)} + \frac12 \bigl(\Lambda_-(1,2) - \Lambda_+ (1,2) \bigr)
= Q(1,2) - \frac12 \, Q_0(1,2) ,
\ee
where we have used the definition \Eqref{bareprop} of $Q_0$.
For simplicity of notation we will refer to both $S(1,2)$ and $Q(1,2)$ as quark propagator,
since they differ only by a kinematic term $Q_0/2$ [\Eqref{bareprop}]. To exhibit the
difference between $S(1,2)$ and $Q(1,2)$ we consider free Dirac fermions ($\bar\gamma=\bar\Gamma_0=0$),
for which from \Eqref{qdse2} follows $Q(1,2) = Q_0(1,2)$ and thus from \Eqref{qprop}
\[
S(1,2)_\mathrm{free\;quarks} = \frac12  \, Q_0(1,2) ,
\]
in agreement with \Eqref{barepropG6}.


\subsection{Gluon Propagator CRDSE}

\begin{figure}
\centering
\includegraphics[height=10ex]{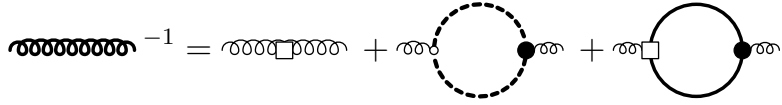}
\caption{\label{fig:gluonDSE}Diagrammatic representation of the CRDSE \eqref{gldse2} for
the gluon propagator. The small empty dot denotes the bare ghost-gluon vertex
$\widetilde\Gamma_0$ [Eq.~\protect\eqref{bggv}], and the thick dashed line stands for the
full (dressed) ghost propagator. The empty square box connected to two (amputated) wavy
lines represents the variational kernel $2\omega$, which equals the inverse bare gluon
propagator.}
\end{figure}

The CRDSE for the static (equal-time) gluon propagator \Eqref{gluonprop}
can be obtained by setting $f=A$ in \Eqref{ans15b}, leading to
\be\label{gldse1}
2 \omega(1,3) \, D(3,2)
- \bar\Gamma_0(4,3;1) \vev*{\xi(3) \, \xi^\dag(4) \, A(2)} - \widetilde\Gamma_0(1;4,3) \vev*{G_A(3,4) \, A(2)} = t(1,2) ,
\ee
where we have used \Eqref{xG6}.
As usual (see Ref.~\cite{Campagnari:2010wc}) the full ghost-gluon vertex $\widetilde\Gamma$ is defined by
[cf.~also the analogous \Eqref{qgv} for the quark-gluon vertex]
\be\label{ggv}
\vev{A(1) G_A(2,3)} = - \widetilde{\Gamma}(2',3';1') \, D(1,1') \, G(2,2') \, G(3',3) ,
\ee
where $G=\langle G_A\rangle$ is the ghost propagator.
Following Refs.~\cite{Feuchter:2004mk,Campagnari:2010wc} we introduce the ghost loop $\chi$ by
\be\label{ghostloop}
2 \chi(1,2) = \widetilde\Gamma_0(3,4;1) \,   G(4,4') \, \widetilde\Gamma(4',3';2) G(3',3) .
\ee
In an analogous way we define the quark loop by
\be\label{quarkloop}
2 \sigma(1,2) = \bar\Gamma_0(3,4;1)  \, Q(4,4') \, \bar\Gamma(4',3';2) \, Q(3',3).
\ee
Introducing furthermore the gluon energy $\Omega$ by
\be\label{G11}
D (1, 2) = \frac{1}{2} \, \Omega^{- 1} (1, 2) ,
\ee
the CRDSE \eqref{gldse1} for the gluon propagator can be cast into the form
\be\label{gldse2}
\Omega(1,2) = \omega(1,2) + \chi(1,2) + \sigma(1,2) ,
\ee
which is diagrammatically represented in Fig.~\ref{fig:gluonDSE}.


\subsection{\label{subsec:qgvdse}Quark-Gluon Vertex CRDSE}

The CRDSEs for the quark and gluon propagator, see Figs.~\ref{fig:fermionDSE} and \ref{fig:gluonDSE},
contain the full quark-gluon vertex $\bar\Gamma$ defined by \Eqref{qgv} and denoted
diagrammatically by a full dot connecting a gluon and two quark
lines. In the DSEs of the functional integral approach in Landau gauge a
substantial dressing of the quark-gluon vertex is required in order to obtain a
sufficient amount of spontaneous breaking of chiral symmetry.\footnote{To date, a complete
solution of the DSE for the quark-gluon vertex has not been obtained, but models for a dressed quark-gluon vertex 
phenomenologically constructed in accord with the Slavnov--Taylor identity exist, see e.g.~Refs.~\cite{Fischer:2008wy,*Fischer:2012vc,Aguilar:2010cn,*Aguilar:2013ac}.}
Although in the present Hamiltonian approach in Coulomb gauge spontaneous breaking of chiral symmetry is 
triggered already by the non-abelian Coulomb interaction \cite{Adler:1984ri,Watson:2011kv,Pak:2011wu,*Pak:2013uba}
[see \Eqref{hamCqq} below] the obtained quark condensate, the corresponding order 
parameter, is far too small \cite{Adler:1984ri}. It increases substantially when the
quark-gluon coupling is included \cite{Pak:2011wu,Pak:2013uba}. However, the quark
condensate obtained in Refs.~\cite{Pak:2011wu,Pak:2013uba} using a bare quark-gluon
vertex is still somewhat too small.
At the moment it is unclear whether the missing strength of chiral symmetry breaking
is due to the use of a bare quark-gluon vertex or due to the approximation for the
propagators.\footnote{%
In Ref.~\cite{Pak:2011wu} the variational approach to QCD was formulated in the usual
second quantization operator formalism, avoiding the introduction of Grassmann fields.
In that formulation it is convenient to take first the fermionic expectation value,
which leaves one with functionals over the gauge fields. In the subsequent gluonic expectation values
in Ref.~\cite{Pak:2011wu} denominators were replaced by their (gluonic) expectation value.
This approximation simplifies the analytical calculation but is unnecessary in
the present CRDSE approach.}
In any case it seems worthwhile to investigate the dressing of the
quark-gluon vertex; this is given by a CRDSE, which we will derive below.

\begin{figure}
\centering
\includegraphics[height=10ex]{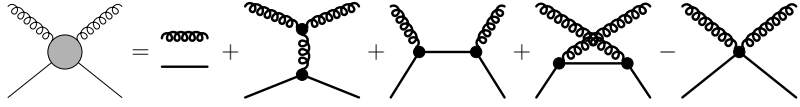}
\caption{\label{fig:vev-2q2A}Vacuum expectation value of two gauge and two Grassmann fields.}
\end{figure}

\begin{figure}
\centering
\includegraphics[height=12ex]{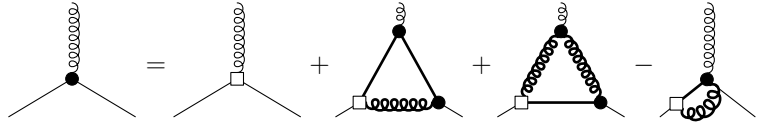}
\caption{\label{fig:qgvDSE}Diagrammatic representation of the DSE \protect\eqref{qgvdse2}
for the quark-gluon vertex.}
\end{figure}

\begin{figure}
\centering
\includegraphics[height=12ex]{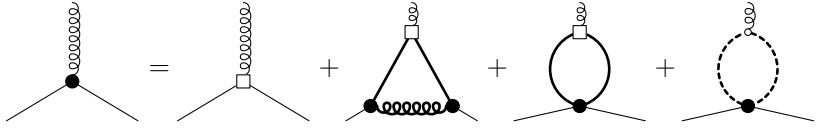}
\caption{\label{fig:qgvDSE2}Alternative CRDSE \protect\eqref{qgvdse3} for the quark-gluon
vertex resulting from the gluonic CRDSE \protect\eqref{ans15b}.}
\end{figure}

Putting $f=\xi^\dag A$ in \Eqref{ans15a} and using $\vev{A}=0$ we have
\be\label{qgvdse1}
\bigl[ Q_0^{-1}(1,4) + \bar\gamma(1,4) \bigr] \vev*{\xi(4) \, \xi^\dag(2) \, A(3) } + \bar\Gamma_0(1,4;5) \vev*{\xi(4) \, A(5) \, \xi^\dag(2) \, A(3)} = 0 .
\ee
The four-point function can be expressed in terms of propagators and vertex
functions in the standard fashion by Legendre transforming the generating functional
of connected Green's functions to the effective action \cite{Campagnari:2010wc}
\[
\vev{\e^{j A + \eta^\dag \xi + \xi^\dag \eta}} \equiv \e^{W[j,\eta^\dag,\eta]} , \qquad
\Gamma[A,\xi,\xi^\dag] + W[j,\eta^\dag,\eta] = j A + \eta^\dag \xi + \xi^\dag \eta .
\]
Taking appropriate derivatives of the effective action one finds for the two-quark-two-gluon
expectation value
\begin{align*}
\langle A(1) \, A(2) \, \xi(3) \, \xi^\dag(4) \rangle = {}&
D(1,2) \, Q(3,4) \nn \\
& + D(1',1) \, D(2',2) \, Q(3,3') \, Q(4',4) \biggl\{
- \bar\Gamma_{\bar{q}qAA}(3',4';1',2')\nn \\
& + \Gamma(1',2',5) D(5,5') \bar{\Gamma}(3',4';5') \nn \\
&+  \bar{\Gamma}(3',5;1') Q(5,5') \bar{\Gamma}(5',4';2') +
\bar{\Gamma}(3',5;2') Q(5,5') \bar{\Gamma}(5',4';1') \Bigr]\biggr\} ,
\end{align*}
where the two-quark-two-gluon proper vertex is defined by
\[
\bar\Gamma_{\bar{q}qAA}(3,4;1,2) =
\frac{\delta^4 \Gamma[A,\xi^\dag,\xi]}
     {\delta \xi(4) \, \delta\xi^\dag(3) \, \delta A(2) \, \delta A(1)}
\biggr|_{A=\xi^\dag=\xi=0} .
\]
This equation is diagrammatically represented in Fig.~\ref{fig:vev-2q2A}.
Inserting this into \Eqref{qgvdse1} the CRDSE for the quark-gluon vertex becomes
\begin{align}\label{qgvdse2}
\bar\Gamma(1,2;3) ={}& \bar\Gamma_0(1,2;3)\nonumber \\
&+ \bar\Gamma_0(1,4;6') Q(4,4') \bar\Gamma(4',5;3) Q(5,5') \bar\Gamma(5',2;6) D(6,6')   \nonumber \\
&+ \bar\Gamma_0(1,4;6') D(4,4') \Gamma(4',5,3) D(5,5') \bar\Gamma(6,2;5') Q(6',6)   \nonumber \\
&- \bar\Gamma_0(1,5;4) D(4,4') Q(5,5') \bar\Gamma_{\bar{q}qAA}(5',2;4',3) .
\end{align}
Equation~\eqref{qgvdse2} is diagrammatically represented in Fig.~\ref{fig:qgvDSE}.
Let us also mention that another CRDSE for the quark-gluon vertex can be obtained by
putting $f=AA$ in the gluonic CRDSE \eqref{ans15b}. This results in
\begin{align}\label{qgvdse3}
\bar\Gamma(1,2;3) ={}& \bar\Gamma_0(1,2;3) \nonumber \\
&+ \bar\Gamma(1,4;6') Q(4,4') \bar\Gamma_0(4',5;3) Q(5,5') \bar\Gamma(5',2;6) D(6,6')   \nonumber \\
&+ \bar\Gamma_{\bar{q}\bar{q}qq}(1,4;2,5) Q(5,5') Q(4',4) \bar\Gamma_0(5',4';3) \nonumber \\
&- \Gamma_{\bar{q}q\bar{c}c}(1,2;4,5) G(5,5') G(4',4) \widetilde\Gamma_0(5',4';3),
\end{align}
which is diagrammatically shown in Fig.~\ref{fig:qgvDSE2}. Its new elements are a
two-ghost-two-fermion vertex $\Gamma_{\bar{q}q\bar{c}c}$ and a four-fermion vertex $\bar\Gamma_{\bar{q}\bar{q}qq}$.
Both equations are equivalent as long as no approximations are introduced.


\subsection{Ghost CRDSE}

\begin{figure}
\centering
\includegraphics[height=8ex]{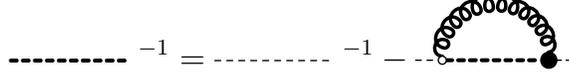}
\caption{\label{fig:ghostDSE1} CRDSE \protect\eqref{ghdse2} for the ghost propagator.}
\end{figure}

For the sake of completeness we quote here also the CRDSE for the ghost propagator, which was 
already derived in Refs.~\cite{Feuchter:2004mk,Campagnari:2010wc}. To obtain this equation
we do not need to explicitly introduce 
ghost fields. Rather, this equation can be obtained already from the operator identity
\[
G_A(1,2) = G_0(1,2) - G_0(1,4) A(3) \, \widetilde{\Gamma}_0(4,5;3) \, G_A(4,2) ,
\]
which follows from the definition of the Faddeev--Popov operator $G^{- 1}_A$ [\Eqref{xyz}]
when this operator is inverted.
Taking the VEV of this identity and using the definition \Eqref{ggv} of the full ghost-gluon vertex
we obtain the ghost CRDSE
\be\label{ghdse2}
G^{-1}(1,2) = G_0^{-1}(1,2) - \widetilde\Gamma_0(1,4;3) D(3,3') G(4,4') \widetilde\Gamma(4',2;3') ,
\ee
which is diagrammatically represented in Fig.~\ref{fig:ghostDSE1}.
The CRDSEs for the higher $n$-point functions of the ghost field can be derived 
by representing the Faddeev-Popov determinant as a functional integral over the ghost fields and employing the standard
DSE techniques, which we are using in the present paper for the quarks and the gluon fields.

The CRDSEs for the gluon and ghost propagators contain also the full ghost-gluon vertex,
see Figs.~\ref{fig:gluonDSE} and \ref{fig:ghostDSE1}. The CRDSE for this vertex was
derived in Ref.~\cite{Campagnari:2010wc} and studied in Ref.~\cite{Campagnari:2011bk}.
It was shown there that its dressing is negligible.

A final comment is in order concerning the CRDSEs derived in this section: For purely Gaussian wave functionals describing independent quasi-particles these CRDSEs
become trivial and are not really necessary, since the higher-order correlation functions
can be entirely expressed in terms of the two-point functions (propagators) by means of
Wick's theorem. However, for interacting theories treated beyond the mean-field approximation
non-Gaussian wave functionals like our vacuum state [Eqs.~\eqref{qcd0} and \eqref{dir5}]
necessarily emerge. Then the CRDSEs derived above allow us to express the various propagators
of QCD in terms of the (so far unknown) variational kernels entering our ansatz for the vacuum wave functional.
In Sect.~\ref{sec:qcdham} we will use these equations to express the vacuum expectation value of the QCD
Hamiltonian in terms of the variational kernels. In this way these CRDSEs enable us to carry out
the variational principle for non-Gaussian wave functionals, which are required for
interacting fields.


\section{\label{sec:qcdham}The QCD Vacuum Energy Density}

With the CRDSEs at hand we are now in a position to express the vacuum expectation value
of the QCD Hamiltonian in terms of the variational kernels. For this purpose we
will first separate the various powers of the fields in the QCD Hamiltonian, so that
their vacuum expectation value results in the various $n$-point functions.

The QCD Hamiltonian in Coulomb gauge is given by \cite{Christ:1980ku}
\begin{multline}\label{X15}
H_{\mathrm{QCD}} =
- \frac12 \int \d[3]x \: \calJ_A^{-1} \frac{\delta}{\delta A_i^a(\vx)} \calJ_A \frac{\delta}{\delta A_i^a(\vx)}
+ \frac12 \int \d[3]x \: \vB^a(\vx) \, \vB^a(\vx) \\
+\int \d[3]x \: \psi^{n\dag}(\vx) \bigl[-\I* \alpha_i \partial_i + \beta m\bigr] \psi^n(\vx)
- g \int \d[3]x \: \psi^{m\dag}(\vx) \alpha_i A_i^a(\vx) \, t^a_{mn} \psi^n(\vx) \\
+ \frac{g^2}{2} \int \d[3]x \d[3]y \: \calJ_A^{-1} \rho^a(\vx) \calJ_A \, F_A^{ab}(\vx,\vy) \, \rho^b(\vy) ,
\end{multline}
where $t^a$ are the generators of $\mathfrak{su}(N)$ in the fundamental representation,
$\calJ_A$ is the Faddeev--Popov determinant [\Eqref{fp}], and
\[
B_{k}^a(\vx) =  \varepsilon_{kij} \left( 
\partial_i A_j^a(\vx) + \frac{g}{2} f^{abc} A_i^b(\vx) \, A_j^c(\vx) \right)
\]
is the chromomagnetic field. Furthermore
\be\label{coulkernel}
F_A^{ab}(\vx,\vy) = \int \d[3]z \: G_A^{ac}(\vx,\vz) \, (-\partial^2_\vz) \, G_A^{cb}(\vz,\vy)
\ee
is the so-called Coulomb kernel, which arises from the resolution of Gauss's law in
Coulomb gauge, and
\[
\rho^a(\vx) = f^{abc} A_i^b(\vx) \frac{\delta}{\I \, \delta A_i^c(\vx)} + \psi^{m\dag}(\vx)t^a_{mn}\psi^n(\vx)
\]
is the colour charge density, which we express as
\[
\rho(1) = \rho_A(1) + \rho_q(1) \eqcolon R(2,3;1) \, A(2) \: \frac{\delta}{\I \, \delta A(3)} + 
\bar{R}(2,3;1) \, \psi^\dag(2) \, \psi(3) .
\]
Here we have introduced the kernels
\begin{subequations}
\begin{align}
R(2,3;1) &= f^{a_1a_2a_3} \, \delta_{i_2i_3} \, \delta(\vx_1-\vx_2) \, \delta(\vx_1-\vx_3) , \\
\label{quarkcharge}
\bar{R}(2,3;1) &= t^{a_1}_{m_2 m_3} \, \delta_{s_2 s_3} \, \delta(\vx_1-\vx_2) \, \delta(\vx_1-\vx_3) .
\end{align}
\end{subequations}

For later convenience we rewrite the total QCD Hamilton operator as
\be\label{qcd2}
H_{\mathrm{QCD}} = H_{\mathrm{YM}} + H_{\mathrm{C}}^{qA} + H_{\mathrm{C}}^{qq} + H_{\mathrm{D}} , \qquad
H_{\mathrm{YM}} = H_E + H_B + H_{\mathrm{C}}^{AA}
\ee
and express the various terms in our compact notation. The kinetic (chromoelectric) part
of the gauge field then reads
\[
H_E = - \frac12 \calJ_A^{-1} \frac{\delta}{\delta A(1)} \calJ_A \frac{\delta}{\delta A(1)} ,
\]
while the chromomagnetic part can be expressed as
\be\label{hamB}
H_B = \frac{1}{2} \int \d[3] x \bigl( B^a_{i} (\vx) \bigr)^2  = - \frac12 A(1) \Delta(1,2) A(2)
+ \frac{1}{3!} T(1,2,3) A(1) A(2) A(3) + \frac{1}{4!} T(1,2,3,4)  A(1) A(2) A(3) A(4) .
\ee
Here we have defined
\be\label{laplacian}
\Delta (1, 2) = \delta^{a_1 a_2} \delta_{i_1i_2} \partial^2_{x_1} \delta (\vx_1-\vx_2).
\ee
Furthermore, the tensor structures $T(1,2,3)$ and $T(1,2,3,4)$ are
irrelevant for the present work and can be found in Ref.~\cite{Campagnari:2010wc}.
The non-abelian Coulomb interaction contains a pure Yang--Mills part
\be\label{hamCAA}
H_{\mathrm{C}}^{AA} = \frac{g^2}{2} \calJ_A^{-1} \rho_A(1) \calJ_A  F_A(1,2) \, \rho_A(2) ,
\ee
a fermion-gluon interaction
\be\label{hamCqA}
H_{\mathrm{C}}^{qA} = \frac{g^2}{2} \left[ \calJ_A^{-1} \rho_A(1) \calJ_A  F_A(1,2) \, \rho_q(2) +  \rho_q(1)  F_A(1,2) \, \rho_A(2) \right] ,
\ee
and a fermionic Coulomb interaction
\be\label{hamCqq}
H_{\mathrm{C}}^{qq} = \frac{g^2}{2} \: \rho_q(1) \, F_A(1,2) \, \rho_q(2) .
\ee
Finally, the one-particle quark Hamiltonian can be written as
\be\label{hamD}
H_{\mathrm{D}} = \psi^\dag(1) \bigl[ h_0(1,2) - J(1,2;3)  \, A(3) \bigr] \psi(2) \equiv
H^{(0)}_{\mathrm{D}} + H^{(1)}_{\mathrm{D}}
\ee
where $h_0$ is defined by \Eqref{h0} and we have introduced the bare quark-gluon vertex
of the QCD Hamiltonian
\be\label{qgcoup}
J(1,2;3) = g \, t^{a_3}_{m_1m_2} \, (\alpha_{i_3})_{s_1s_2} \, \delta(\vx_1-\vx_2) \, \delta(\vx_1-\vx_3) .
\ee

To carry out the non-perturbative variational approach
we now evaluate the expectation value of the QCD Hamiltonian \Eqref{qcd2} in
the state defined by Eqs.~\eqref{qcd0}, \eqref{ans11}, and \eqref{dir5}. We carry out this
evaluation up to two-loop level, so that the corresponding equations of motion following
from the variation of the energy will contain at most one loop. Let us emphasize, however,
that loops are here defined in terms of the non-perturbative propagators and vertices.

The magnetic term $H_B$ [\Eqref{hamB}] is insensitive to the quark part of the vacuum wave 
functional and hence yields the same contribution as in the pure Yang--Mills theory.
Furthermore, the quark contribution to $\langle H_{\mathrm{C}}^{AA} \rangle$ [\Eqref{hamCAA}]
contains more than two loops, which we do not include here.  Also the fermion-gluon
Coulomb interaction $H^{qA}_{\mathrm{C}}$ [\Eqref{hamCqA}] yields contributions only
beyond two loops and is hence discarded.

When a Gaussian functional is used for the Yang--Mills sector [see Eqs.~\eqref{qcd0} and
\eqref{G5}] the cubic term of the magnetic energy \Eqref{hamB} does not contribute.
Furthermore, the quartic term gives rise to a tadpole in the gluonic gap equation
\cite{Feuchter:2004mk}, which can be absorbed into a renormalization constant. Note also
that the contribution of the quartic term to the energy vanishes in dimensional
regularization. Therefore in the following we will omit the cubic and quartic term of
the magnetic energy \Eqref{hamB}, which then reduces to
\be\label{G16}
\vev{H_B} = -\frac12 \Delta(1,2) D(2,1)
\ee
where $D(2,1)$ is the gluon propagator \Eqref{gluonprop}.

Due to overall translational invariance, the vacuum expectation value of the various
terms of the Hamiltonian \Eqref{X15} always contains a diverging factor $(2\pi)^3 \delta(\vp=0)$,
which is nothing but the spatial volume $V$. This factor disappears when the energy density
\[
e=\vev*{H}/V
\]
is considered.


\subsection{Single-Particle Hamiltonian}

\begin{figure}
\centering
\includegraphics[width=.1\linewidth]{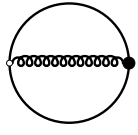}
\caption{\label{fig:en_qAq}Diagrammatic representation of the last term in the r.h.s.~of~\protect\Eqref{G16A}.}
\end{figure}

The vacuum expectation value of the single-particle quark Hamiltonian \Eqref{hamD} is easily evaluated
by means of \Eqref{vev3a}
\[
\begin{split}
\vev{H_\mathrm{D}} &= h_0(1,2) \vev*{\psi^\dag(1) \psi(2)} - J(1,2;3) \vev*{\psi^\dag(1) \psi(2) \, A(3)} \\
&= h_0(1,2) \bigl[ \vev*{\xi^\dag(1) \xi(2)} + \Lambda_+(2,1)\bigr] - J(1,2;3) \bigl[ \vev*{\xi^\dag(1) \xi(2) \, A(3)} + \Lambda_+(2,1) \vev{A(3)}\bigr] .
\end{split}
\]
Using $\vev{A}=0$ and the definition \Eqref{qgv} of the full quark-gluon vertex $\bar\Gamma$
the above  expression can be cast into the form
\be\label{G16A}
\vev{H_\mathrm{D}} = - h_0(1,2) [ Q(2,1) - \Lambda_+(2,1)] - J(1,2;3) D(3,3') Q(2,2') \bar\Gamma(2',1';3') Q(1',1).
\ee
Here the last term arises from the direct coupling of the quarks to the gluons through
the bare vertex $J(1,2;3)$ [\Eqref{qgcoup}] in the QCD Hamiltonian. This term is diagrammatically
represented in Fig.~\ref{fig:en_qAq}.
Besides the bare vertex $J$ it contains also the full quark-gluon vertex $\bar\Gamma$,
whose CRDSEs have been derived in Sect.~\ref{subsec:qgvdse} [Eqs.~\eqref{qgvdse2} and \eqref{qgvdse3}].
These relate the full quark-gluon vertex via \Eqref{ans9} to the fermionic variational kernels.


\subsection{Fermion-Fermion Coulomb Interaction}

For the spontaneous breaking of chiral symmetry (SB$\chi$S) the quark part of the Coulomb interaction 
$H^{qq}_\mathrm{C}$ [\Eqref{hamCqq}] seems to be crucial.
This term alone triggers already SB$\chi$S \cite{Adler:1984ri}, albeit not of sufficient strength.
On the other hand it was shown in Refs.~\cite{Pak:2011wu,Pak:2013uba}
that the quark-gluon coupling (in $H_\mathrm{D}$) alone does not provide SB$\chi$S,
at least within the approximation used in Refs.~\cite{Pak:2011wu,Pak:2013uba}.\footnote{%
In Refs.~\cite{Pak:2011wu,Pak:2013uba} no dressing of the (variational) quark-gluon vertex
as described by the CRDSE~\eqref{qgvdse2} was included. It is entirely possible that when
the full dressing of the quark-gluon vertex is included SB$\chi$S does take place without
including the Coulomb term $H_\mathrm{C}^{qq}$. In fact the results of recent lattice
investigations \cite{Glozman:2012fj} could be interpreted in favour of such a scenario \cite{Glozman:2015qva}.} 
However, the quark-gluon coupling
in $H_\mathrm{D}$ substantially increases the amount of chiral symmetry breaking once
$H_\mathrm{C}^{qq}$ is included \cite{Pak:2011wu,Pak:2013uba}.

The expectation value of the quark Coulomb interaction \Eqref{hamCqq}
\begin{align}
E_\mathrm{C}^{qq} \equiv \vev{H_\mathrm{C}^{qq}} &= \frac{g^2}{2} \bar{R}(3,4;1) \bar{R}(5,6;2)
\vev{\psi^\dag(3) \,\psi(4) \, F_A(1,2) \, \psi^\dag(5) \,\psi(6)} \nn \\
\intertext{is taken by means of \Eqref{vev4}, which yields}
E_\mathrm{C}^{qq}&=\frac{g^2}{2} \bar{R}(3,4;1) \bar{R}(5,6;2) \Bigl\{
\begin{aligned}[t]
&\vev{\xi^\dag(3) \, \xi(4) \, F_A(1,2) \, \xi^\dag(5) \, \xi(6)} \\
&+\vev{\xi^\dag(3) \xi(4) F_A(1,2)} \Lambda_+(6,5) + \vev{\xi(4) \xi^\dag(5) F_A(1,2)} \Lambda_+(6,3) \\
&+\vev{\xi^\dag(5) \xi(6) F_A(1,2)}\Lambda_+(4,3) + \vev{\xi^\dag(3) \xi(6) F_A(1,2)}\Lambda_-(4,5)\\
&+\bigl[ \Lambda_+(4,3)\Lambda_+(6,5) + \Lambda_+(6,3)\Lambda_-(4,5) \bigr] \vev{F_A(1,2)} \Bigr\} .
\end{aligned}
\raisetag{2ex}
\label{eCqq1}
\end{align}
Up to two loops in the energy we can replace the Coulomb kernel $F_A$ [\Eqref{coulkernel}]
by its (gluonic) vacuum expectation value $F\equiv\vev*{F_A}$. Furthermore, since the
Dirac projectors $\Lambda_\pm$ are the unit matrix in colour space their contraction
with the kernels $\bar{R}$ [\Eqref{quarkcharge}] of the quark colour charge density
results in the trace of the generators of the gauge group, which vanishes.\footnote{%
This would not be the case within an abelian theory. The arising singular terms can
nevertheless be eliminated by an appropriate redefinition of the charge operator
$\psi^\dag\psi \to \frac12 \comm{\psi^\dag}{\psi}$.} 
For this reason the second, fourth, and sixth term in the brackets
on the r.h.s.~of \Eqref{eCqq1} vanish and we are left with
\be\label{eCqq3}
E_\mathrm{C}^{qq} \simeq \frac{g^2}{2} \bar{R}(3,4;1) \bar{R}(5,6;2) F(1,2)
\Bigl[ \vev*{\xi(4) \,\xi^\dag(3) \, \xi(6) \, \xi^\dag(5)} + Q(4,5) \, \Lambda_+(6,3) - Q(6,3) \, \Lambda_-(4,5)  + \Lambda_+(6,3) \, \Lambda_-(4,5)\Bigr].
\ee
Finally, up to two-loop order in the energy it is sufficient to take the lowest order
contribution to the fermion four-point function
\be\label{eCqq4}
\vev*{\xi(4) \,\xi^\dag(3) \, \xi(6) \, \xi^\dag(5)} = Q(4,3) \, Q(6,5) - Q(4,5) \, Q(6,3) + \text{connected terms.}
\ee
Since the quark propagator $Q(1,2)$ is colour diagonal, when \Eqref{eCqq4} is inserted
into \Eqref{eCqq3} the first term on the right-hand side of \Eqref{eCqq4} gives also rise
to a trace over the colour generators and thus to a vanishing contribution to the quark Coulomb energy
\Eqref{eCqq3}, which then becomes
\be\label{G18}
E_\mathrm{C}^{qq} \simeq - \frac{g^2}{2} \bar{R}(3,4;1) \bar{R}(5,6;2) F(1,2) \, \bigl[ Q(4,5) + \Lambda_-(4,5) \bigr]  \bigl[ Q(6,3) - \Lambda_+(6,3) \bigr] .
\ee


\subsection{The Chromoelectric Energy}

Contrary to the magnetic energy $\vev*{H_B}$ and the gluonic Coulomb energy $\vev*{H_{\mathrm{C}}^{AA}}$,
the chromoelectric energy $\vev*{H_E}$ does receive additional contributions from the
quark sector at the considered two-loop order due to the quark-gluon coupling \Eqref{ans2} in
the fermionic wave functional \Eqref{dir5}. With the explicit form of the vacuum wave functional
we find from \Eqref{qcd1} after an integration by parts with respect to the gluon field
\begin{align}
\vev*{H_E} &= \frac12 \int\calD A \, \calD\xi^\dag \, \calD\xi \: \e^{-\mu}
\left[ \frac{\delta}{\delta A(1)} \: \e^{-S_f^*-\frac12 S_A}\right] \calJ_A \frac{\delta}{\delta A(1)} \e^{-S_f-\frac12 S_A} \nn\\
&= \frac12
\vev*{ \left[ \frac{\delta S_f^*}{\delta A(1)} + \frac12 \frac{\delta S_A}{\delta A(1)} \right]
      \left[ \frac{\delta S_f}{\delta A(1)} + \frac12 \frac{\delta S_A}{\delta A(1)} \right] } \nn\\
&= \frac18 \vev*{\frac{\delta S_A}{\delta A(1)} \frac{\delta S_A}{\delta A(1)}}
+ \frac14 \vev*{\frac{\delta S_A}{\delta A(1)} \frac{\delta (S_f+S_f^*)}{\delta A(1)} }
+ \frac12 \vev*{ \frac{\delta S_f^*}{\delta A(1)} \frac{\delta S_f}{\delta A(1)} } .
\label{eE1}
\end{align}
To work out the remaining vacuum expectation values we use here the CRDSEs derived in 
Sec.~\ref{subsec:qcd}. For this purpose, using $S=S_A+S_f+S_f^*+\mu$ [see \Eqref{vac7}] and $\delta\mu/\delta A=0$,
we rewrite the general CRDSE \eqref{dses6} as
\be\label{dsegen}
\vev*{\frac{\delta S_A}{\delta A(1)} \: f}
= \vev*{\frac{\delta f}{\delta A(1)}} - \vev*{\frac{\delta (S_f+S_f^*)}{\delta A(1)} \: f} + \widetilde{\Gamma}_0(3,2;1) \vev{G_A(2,3) f} .
\ee
For the first two terms in \Eqref{eE1} we use this CRDSE with $f = \delta S_A / \delta A$ and obtain
\begin{align}
& \vev*{\frac{\delta S_A}{\delta A(1)} \frac{\delta S_A}{\delta A(1)}}
+ 2 \vev*{\frac{\delta S_A}{\delta A(1)} \frac{\delta (S_f+S_f^*)}{\delta A(1)} } \nn\\
&= \vev*{\frac{\delta^2 S_A}{\delta A(1) \, \delta A(1)}}
+ \widetilde{\Gamma}_0(3,2;1) \vev*{G_A(2,3)\frac{\delta S_A}{\delta A(1)}}
+\vev*{\frac{\delta (S_f+S_f^*)}{\delta A(1)} \frac{\delta S_A}{\delta A(1)}} .
\label{eE2}
\end{align}
For the last term we can again use \Eqref{dsegen} putting $f=\delta(S_f+S_f^*)/\delta A$,
yielding
\[
\vev*{\frac{\delta (S_f+S_f^*)}{\delta A(1)} \frac{\delta S_A}{\delta A(1)}}
=\vev*{\frac{\delta^2 (S_f+S_f^*)}{\delta A(1) \, \delta A(1)}}
+ \widetilde{\Gamma}_0(3,2;1) \vev*{G_A(2,3)\frac{\delta (S_f+S_f^*)}{\delta A(1)}}
-\vev*{\frac{\delta (S_f+S_f^*)}{\delta A(1)} \frac{\delta (S_f+S_f^*)}{\delta A(1)}} .
\]
Inserting this expression into \Eqref{eE2} we obtain
\begin{multline*}
\vev*{\frac{\delta S_A}{\delta A(1)} \frac{\delta S_A}{\delta A(1)}} + 2 \vev*{\frac{\delta S_A}{\delta A(1)} \frac{\delta (S_f+S_f^*)}{\delta A(1)}} \\
= \vev*{\frac{\delta^2 S}{\delta A(1) \, \delta A(1)}}  
+ \widetilde{\Gamma}_0(3,2;1) \vev*{G_A(2,3)\frac{\delta S}{\delta A(1)}}
-\vev*{\frac{\delta (S_f+S_f^*)}{\delta A(1)} \frac{\delta (S_f+S_f^*)}{\delta A(1)}} .
\end{multline*}
Using the above derived expressions we can finally write the Yang--Mills chromoelectric energy \Eqref{eE1} as
\be\label{eE17}
\vev{H_E} = \frac{1}{8} \vev*{\frac{\delta^2 S}{\delta A(1) \, \delta A(1)}}
- \frac18 \vev*{\frac{\delta(S_f-S_f^*)}{\delta A(1)} \frac{\delta(S_f-S_f^*)}{\delta A(1)}}
+ \frac18 \widetilde{\Gamma}_0(3,2;1) \vev*{G_A (2, 3) \frac{\delta S}{\delta A (1)} } .
\ee
Equation~\eqref{eE17} is, so far, exact.
Restricting ourselves to the Gaussian ansatz for the gluonic part of vacuum wave functional  
[see \Eqref{ans11}] the various terms can be explicitly calculated. In the last term 
the definition of the ghost-gluon vertex [\Eqref{ggv}] has to be used. One finds then for the 
chromoelectric energy
\be\label{eE7}
\vev*{H_E} = E_E^\mathrm{YM} + E_E^\mathrm{Q} ,
\ee
where
\be\label{eEym}
 E_E^\mathrm{YM} = \frac{1}{4} \bigl[ \Omega(1,2) - \chi(1,2) \bigr] \Omega^{-1}(2,3) \bigl[ \Omega(3,1) - \chi(3,1) \bigr]
\ee
is the contribution arising from the pure Yang--Mills sector, and
\be\label{eEq}
E_E^\mathrm{Q} = - \frac14 \: \sigma(1,1) + \frac14 \: \sigma_-(1,1)
\ee
is the explicit contribution of the quarks to the kinetic energy of the gluons. In \Eqref{eEq}
$\sigma_-$ is defined analogously to $\sigma$ [\Eqref{quarkloop}], however, with the bare and 
full quark gluon vertex, $\bar{\Gamma}_0$ and $\bar{\Gamma}$, both replaced by
\be\label{nG19}
\bar\Gamma_-(1,2;3) = \Lambda_{+}(1,1') \, K(1',2';3) \, \Lambda_{-}(2',2) - \Lambda_{-}(1,1') \, K^\dag(1',2;3) \, \Lambda_{+}(2',2) .
\ee
[This follows from the terms $S_f-S_f^*$ in \Eqref{eE17}.]
Using the properties \Eqref{1447X1B} of the projectors $\Lambda_\pm$ and \Eqref{bareprop},
the quantity $\bar\Gamma_-$ [\Eqref{nG19}] can be written as
\be\label{186x1}
\bar\Gamma_-(1,2;3) = Q_0(1,1') \, \bar\Gamma_0(1',2;3) = - \bar\Gamma_0(1,2';3) \, Q_0(2',2) .
\ee


\subsection{\label{sec:toten}The Total Energy}

For carrying out the variation of the energy let us summarize the various energy contributions.
To the order of approximation considered in the present work (two loops in the energy)
the total energy is given by [cf.~\Eqref{qcd2}]
\be\label{G19}
E_\mathrm{QCD} \equiv \vev{H_\mathrm{QCD}} \simeq \vev{H_\mathrm{YM}} + \vev{H_\mathrm{D}} + \vev{H_\mathrm{C}^{qq}} ,
\ee
with
\[
\vev{H_\mathrm{YM}} = \vev{H_E} + \vev{H_B} + \vev{H_\mathrm{C}^{AA}} .
\]
Here $\vev{H_B}$, $\vev{H_\mathrm{D}}$, $\vev{H_\mathrm{C}^{qq}}$, and $\vev{H_E}$ are
given by Eqs.~\eqref{G16}, \eqref{G16A}, \eqref{G18}, and \eqref{eE7}, respectively.
Furthermore, the expression for $\vev{H_\mathrm{C}^{AA}}$ was given in Ref.~\cite{Feuchter:2004mk}.
As shown in Ref.~\cite{Heffner:2012sx}, on a quantitative level this quantity is
completely irrelevant and will hence be ignored in the following. For subsequent considerations
it will be convenient to rewrite the energy \Eqref{G19} in the form
\[
E_\mathrm{QCD} = E_\mathrm{YM} + E_\mathrm{Q} \, ,
\]
where
\[
E_\mathrm{YM} = E_E^\mathrm{YM} + \vev{H_B}
\]
is the energy of the Yang--Mills sector, for which we get from Eqs.~\eqref{G16} and \eqref{eEym}
\[
E_\mathrm{YM} = \frac12 \Bigl[\frac12 D^{-1}(1,2) - \chi(1,2) \Bigr] D(2,3) \Bigl[\frac12 D^{-1}(3,1) - \chi(3,1) \Bigr] - \frac12 \Delta(1,2) D(2,1) .
\]
The energy of the quarks is given by
\be\label{1447X21A}
E_\mathrm{Q} = \vev{H_\mathrm{D}} + E_\mathrm{C}^{qq} + E_E^\mathrm{Q}
\ee
where the Dirac energy $\vev{H_\mathrm{D}} = E_\mathrm{D}^{(0)} + E_\mathrm{D}^{(1)}$
was given in \Eqref{G16A}. Furthermore, the quark energy \Eqref{1447X21A} includes
the quark contribution to the chromoelectric energy $E_E^\mathrm{Q}$ [\Eqref{eEq}] as
well as the non-abelian Coulomb interaction of the quarks [\Eqref{G18}].

The expressions given above for the quark energy can be substantially simplified by
noticing that they consist of linear chains of fermionic matrices which are
connected by ordinary matrix multiplication. Therefore, without loss of information we
can skip the fermionic indices (but keep the bosonic ones) and assume ordinary matrix
multiplication for the fermionic objects. This we will do in the rest of this paper's
body. The quark energy \Eqref{1447X21A} is then given by
\begin{align}\label{lastone}
E_\mathrm{Q} ={}& - \Tr\bigl[h_0 \bigl(Q-\tfrac12 Q_0\bigr)\bigr]
- D(1,2) \Tr \bigl[ J(1) Q \bar\Gamma(2) Q \bigr] \nn\\
&- \frac18 \Tr \bigl[ \bar\Gamma_0(1) Q \bar\Gamma(1) Q \bigr] - \frac18 \Tr \bigl[ Q_0 \bar\Gamma_0(1) Q \bar\Gamma_0(1) Q_0 Q \bigr] \nn\\
&- \frac{g^2}{2} \, F(1,2) \Tr \bigl[ \bar{R}(1) \bigl(Q-\tfrac12 Q_0\bigr) \bar{R}(2) \bigl(Q-\tfrac12 Q_0\bigr) -\tfrac14 \bar{R}(1)\bar{R}(2) \bigr] ,
\end{align}
the trace being over fermionic indices only.

Above we have succeeded to express the vacuum expectation value of the QCD Hamiltonian
in terms of the variational kernels (denoted graphically by open square boxes) and the
various $n$-point functions (denoted graphically by full dots), the latter being themselves
functionals of the variational kernels through the CRDSEs. In addition, the energy contains
the bare vertices of the QCD Hamiltonian, denoted graphically by open circles.

We are now in a position to carry out the variation of the energy. This will result in a set of gap equations,
which have to be solved together with the CRDSEs.


\section{\label{sec:var_prin}The Variational Principle}

Our trial wave functional [see Eqs.~\eqref{qcd0}, \eqref{ans11}--\eqref{ans2}] contains three variational kernels: $\omega$ of the
Yang--Mills wave functional, and $K_0$ and $K$ of the quark wave functional. In carrying
out the variations with respect to these kernels we will ignore implicit dependences
which will give rise to higher-order loops in the resulting gap equations.
This implies in particular that we will ignore
the dependence of the ghost propagator (and hence of $\chi$) on the gluon kernel,
as we did already previously in the treatment of the Yang--Mills sector \cite{Feuchter:2004mk}.
In the same spirit we will ignore
the dependence of the gluon propagator on the fermionic kernels $K_0$ and $K$
as well as the implicit dependence of the quark propagator on $\omega$
with the exception of the free single-particle energy $E_\mathrm{D}^{(0)}$ [\Eqref{G16A}], where we
will include the dependence of $Q$ on the gluon propagator, since this contributes
only a one-loop term to the gap equation.
The explicit derivation of the gap equations is given in App.~\ref{sec:app_ge}.

\begin{figure}
\centering
\includegraphics[width=.6\linewidth]{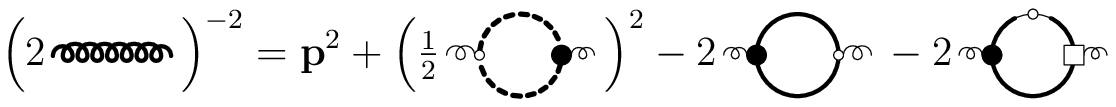}
\caption{\label{fig:gluongap}Diagrammatic representation of \protect\Eqref{G4A}. Small
open circles represent the ``vertices'' occurring in the Hamiltonian.}
\end{figure}

The variational equation with respect to $\omega$ can be combined with the CRDSE~\eqref{gldse2}
as explained in App.~\ref{sec:app_ge}, resulting in
\be\label{G4A}
\Omega^2(1,2) = -\Delta(1,2) + \chi^2(1,2)
- 2 \Tr \bigl[ \bar\Gamma(1) \, Q \, J(2) \, Q \bigr]
-2 \Tr \bigl[\bar\Gamma(1) \, Q \, h_0 Q \bar\Gamma_0(2) Q \bigr].
\ee
Here the trace is over the fermionic indices only, as it should be clear from the context.
Furthermore, $\Delta(1,2)$ is the Laplacian \Eqref{laplacian}, $\chi$ is the ghost loop
[\Eqref{ghostloop}], $Q$ is the full quark propagator [\Eqref{grassprop}], $J$ [\Eqref{qgcoup}]
is the quark-gluon coupling of the QCD Hamiltonian, and $\bar\Gamma_0$ [\Eqref{ans9}] is
the quark-gluon variational kernel of our trial wave functional \Eqref{qcd0}. Finally,
$\bar\Gamma$ is the corresponding dressed quark-gluon vertex \Eqref{qgv}, which is related to the
bare one $\bar\Gamma_0$ by the CRDSE~\eqref{qgvdse2} [or \Eqref{qgvdse3}].
Equation~\eqref{G4A} generalizes the gluonic gap equation obtained in Refs.~\cite{Feuchter:2004mk,Campagnari:2010wc} to full QCD\footnote{%
In Refs.~\cite{Feuchter:2004mk,Campagnari:2010wc} the gluonic Coulomb term $H_\mathrm{C}^{AA}$ [\Eqref{hamCAA}]
was also included, which results in additional terms in the gap equation.} 
and is diagrammatically represented in Fig.~\ref{fig:gluongap}.
Note that the quarks contribute threefold to this equation: 
i) through the last but one term, which is a quark loop arising from the quark-gluon coupling
in the QCD Hamiltonian, 
ii) through the last term, which is a quark loop arising from the free quark energy due to the dependence
of the quark propagator on the gluon propagator [see \Eqref{G20} below], and 
iii) through the quark loop $\sigma$ [\Eqref{quarkloop}] entering the gluon CRDSE \eqref{gldse2}
for $\Omega$. The latter arises entirely from the quark-gluon coupling $\bar\Gamma_0$ in the QCD wave functional.
In App.~\ref{sec:app_en} the gluon gap equation \eqref{G4A} is used to simplify the
expression for the stationary energy of the QCD vacuum, which will be needed in future
investigations.

\begin{figure}
\centering
\includegraphics[width=.8\linewidth]{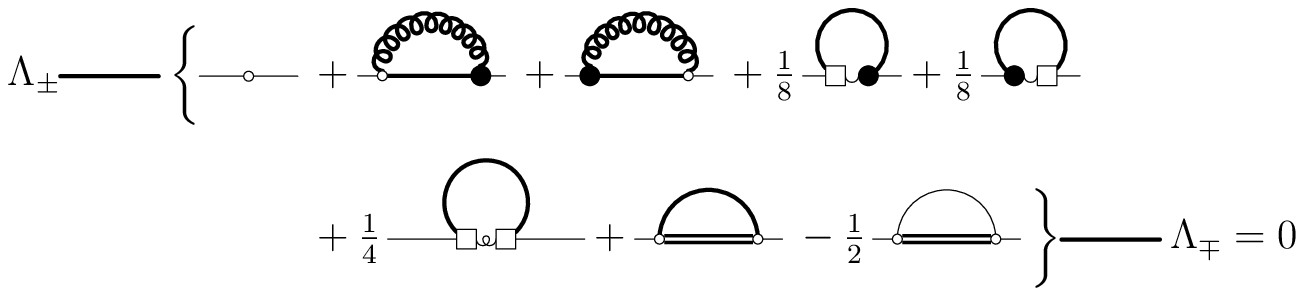}
\caption{\label{fig:quarkgap}Diagrammatic representation of the quark equation resulting
from Eqs.~\protect\eqref{G6Baaa} and \protect\eqref{G6Bxxx}. The double line represents the
Coulomb propagator $F\equiv\langle F_A \rangle$ [Eq.~\protect\eqref{coulkernel}].}
\end{figure}

The variation with respect to the biquark kernel $K_0$ leads to the conditions
\be\label{G6Baaa}
\Lambda_- Q h Q \Lambda_+ = 0 , \qquad
\Lambda_+ Q h Q \Lambda_- = 0 ,
\ee
where $h=-\delta E_\mathrm{QCD}/\delta Q$ is an effective single-quark Hamiltonian
\begin{align}\label{G6Bxxx}
h = h_0 &+  D(1,2) \bigl[ J(1) Q \bar\Gamma(2) + \bar\Gamma(1) Q J(2) \bigr] \nonumber \\
&+\frac18 \Bigl\{ \bar\Gamma_0(1) Q \bar\Gamma(1) + \bar\Gamma(1) Q \bar\Gamma_0(1) \Bigr\}
+ \frac14 \,Q_0 \bar\Gamma_0(1) Q \bar\Gamma_0(1) Q_0 \nonumber \\
&+ g^2 F(1,2) \bar{R}(1) \bigl[Q - \tfrac12 Q_0 \bigr] \bar{R}(2) .
\end{align}
The quark gap equations are shown diagrammatically in Fig.~\ref{fig:quarkgap}.
In the effective single-particle Hamiltonian, $h_0$ is the Dirac Hamiltonian of
free fermions while the remaining terms on the r.h.s.\ have all the same structure: the
quark propagator $Q$ (or its free counterpart $Q_0$) is sandwiched by quark-gluon
couplings: $J$ is the quark-gluon coupling in the QCD Dirac Hamiltonian, $\bar\Gamma_0$
and $\bar\Gamma$ are, respectively, bare and dressed quark-gluon kernels of our QCD
wave functional, and $\bar{R}$ [\Eqref{quarkcharge}] is the coupling vertex of the quarks to the Coulomb kernel.

\begin{figure}
\centering
\includegraphics[width=.6\linewidth]{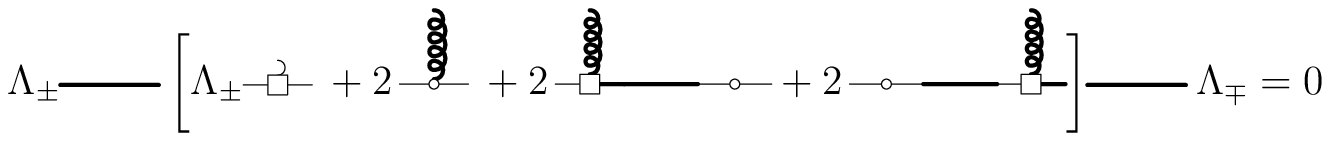}
\caption{\label{fig:vecgap}Diagrammatic representation of the gap equation \protect\eqref{vecgap}
for the vector kernel.}
\end{figure}

The variation of the energy with respect to the vector kernel $K$ or $K^\dag$ is carried
out in App.~\ref{app:vkg}. Thereby the CRDSE~\eqref{qgvdse2} is used to find the variation
of the full (dressed) quark-gluon vertex with respect to the kernels $K$, $K^\dag$. For
simplicity, we quote here the resulting variational equations for $K$, $K^\dag$ only in
the bare-vertex approximation [\Eqref{ststst}]
\be\label{vecgap}
\Lambda_\pm \, Q \bigl\{ \Lambda_\pm \bar\Gamma_0(1) + 2 D(1,2) \bigl[ J(2) + h_0 Q \bar\Gamma_0(2)  + \bar\Gamma_0(2) Q h_0 \bigr] \bigr\} Q \Lambda_\mp = 0,
\ee
which is represented diagrammatically in Fig.~\ref{fig:vecgap}.
The kernel $K$ enters here through the bare quark-gluon vertex $\bar\Gamma_0$ [\Eqref{ans9}].
In fact, \Eqref{vecgap} can be explicitly solved for $K$ \cite{Campagnari:tbp}.

Equations \eqref{G4A}, \eqref{G6Bxxx}, and \eqref{vecgap} provide the gap equations for the
variational kernels of our trial ansatz given by Eqs.~\eqref{qcd0}, \eqref{ans11} and \eqref{dir5} for the QCD vacuum wave functional. These equations have to be solved
together with the CRDSEs for the various propagators and vertices occurring in the variational equations.
In a first step one will use the bare vertex approximation, which results in an explicit
expression for the vector kernel $K(1,2;3)$ in terms of the quark and
gluon propagators. Furthermore, one will do a quenched calculation using the gluon propagator
(known from the present approach to the Yang--Mills sector and also from the lattice calculation)
as input for the quark sector. Such calculations are presently carried out.


\section{\label{sec:conc}Conclusions}

The variational approach to the Hamiltonian formulation of interacting quantum field
theories proposed in Ref.~\cite{Campagnari:2010wc} and developed there for Yang--Mills
theory was extended to full QCD. The main feature of this approach is the use of CRDSEs
to express the vacuum expectation values of powers of field operators (i.e.,~$n$-point
functions) in terms of the variational kernels occurring in the exponent of the vacuum
wave functional. In this way the variational approach can be carried out for non-Gaussian
wave functionals, i.e.,~for interacting quantum field theories. To make use of the standard
DSE techniques this approach requires the use of the coherent fermion state basis
of Fock space, which is expressed in terms of Grassmann variables.

Assuming a vacuum wave functional which contains the coupling of the quarks to the
gluons we have derived the necessary CRDSEs. By means of these CRDSEs we have expressed
the vacuum expectation value of the QCD-Hamiltonian in Coulomb gauge in terms of the
variational kernels of the wave functionals and carried out the variation of the energy,
resulting in a set of so-called gap equations. These  gap equations have to be solved
together with the pertinent CRDSEs.

At first sight it seems that the present variational approach is quite cumbersome and
less economic than the conventional DSE approach in the functional integral formulation
of QCD in Landau gauge \cite{Alkofer:2000wg,Fischer:2008uz,Binosi:2009qm}. There one has
to solve the standard DSEs where the bare vertices are defined by the classical action
of QCD. In the present approach we have to solve the CRDSEs, which are structurally similar
to (and at least as complicated as) the usual DSEs. Moreover, contrary to the
usual DSEs the bare vertices in the CRDSEs are not known a priori but are variational
kernels, which have to be found by solving the gap equations. So it seems
that our variational approach is much more expensive than the conventional
DSE approach to QCD in Landau gauge. However, the infinite tower of DSEs has to be truncated
for practical reasons and there is usually little control over the quality of the approximation
achieved. Also in our approach we have to truncate the tower of CRDSEs. However, whatever
truncation we use, the variational principle (i.e.,~the gap equations) will provide us with the
optimal choice of bare vertices for that truncation. We can therefore expect that the
``bare'' vertices of the CRDSEs, i.e.,~the variational kernels, capture some of the physics
lost by the corresponding truncation of the usual DSEs. In fact our ``bare'' vertices obtained
by solving the gap equations are not at all ``bare'' but resemble more dressed
vertices of the usual DSE approach \cite{Campagnari:tbp}. As an illustrative example
consider the quark-gluon vertex. In the conventional DSE approach in Landau gauge no chiral
symmetry breaking is obtained when a bare quark-gluon vertex is used in the quark DSE.
In our approach we do get chiral symmetry breaking even in the bare-vertex approximation.

In the future we plan to use the approach developed in this paper for a realistic
description of the spontaneous breaking of chiral symmetry in the QCD vacuum. Furthermore,
we intend to extend this approach to QCD at finite temperature and finite baryon density.

The present approach is quite general and in principle can be applied to any interacting
quantum field theory as well as to interacting many-body systems.

\begin{acknowledgments}
The authors thank P.~Watson and M.~Quandt for a critical reading of the manuscript and
useful comments. This work was supported by the Deutsche Forschungsgemeinschaft (DFG)
under contract No.~Re856/10-1.
\end{acknowledgments}


\appendix
\section{Diagrammatics}
\allowdisplaybreaks\noindent
Propagators:
\begin{alignat*}{3}
\parbox{4.1em}{\includegraphics[width=4em]{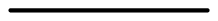}} \quad&\equiv\; Q(1,2)\qquad &&\text{\;[\Eqref{grassprop}]}
\\
\parbox{4.1em}{\includegraphics[width=4em]{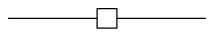}} \quad&\equiv\; \bar\gamma(1,2) &&\text{\;[\Eqref{ans7}]}
\\
\parbox{4.1em}{\includegraphics[width=4em]{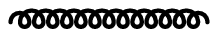}} \quad&\equiv\; D(1,2) &&\text{\;[\Eqref{gluonprop}]}
\\
\parbox{4.1em}{\includegraphics[width=4em]{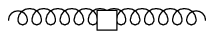}} \quad&\equiv\; 2\omega(1,2) &&\text{\;[\Eqref{ans11}]}
\\
\parbox{4.1em}{\includegraphics[width=4em]{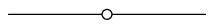}} \quad&\equiv\; h_0(1,2) &&\text{\;[\Eqref{h0}]}
\\
\parbox{4.1em}{\includegraphics[width=4em]{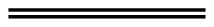}} \quad&\equiv\; \langle F_A(1,2)\rangle &&\text{\;[\Eqref{coulkernel}]} 
\end{alignat*}
Vertices:
\begin{alignat*}{6}
\parbox{4.1em}{\includegraphics[width=4em]{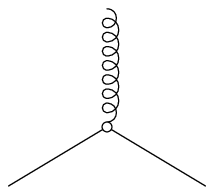}} \quad&\equiv\; J(1,2;3) \quad&&\text{\;[\Eqref{qgcoup}]}
&\qquad\qquad
\parbox{4.1em}{\includegraphics[width=4em]{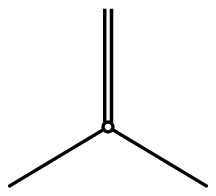}} \quad&\equiv\; \bar{R}(1,2;3) \quad&&\text{\;[\Eqref{quarkcharge}]}
\\
\parbox{4.1em}{\includegraphics[width=4em]{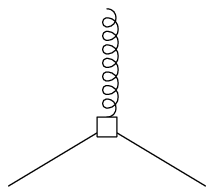}} \quad&\equiv\; \bar\Gamma_0(1,2;3) &&\text{\;[\Eqref{ans9}]}
&
\parbox{4.1em}{\includegraphics[width=4em]{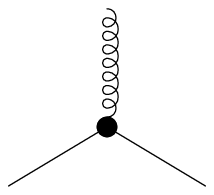}} \quad&\equiv\; \bar\Gamma(1,2;3) &&\text{\;[\Eqref{qgv}]}
\\
\parbox{4.1em}{\includegraphics[width=4em]{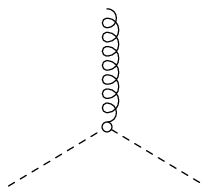}} \quad&\equiv\; \widetilde\Gamma_0(1,2;3) &&\text{\;[\Eqref{bggv}]}
&
\parbox{4.1em}{\includegraphics[width=4em]{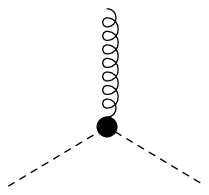}} \quad&\equiv\; \widetilde\Gamma(1,2;3) &&\text{\;[\Eqref{ggv}]}
\end{alignat*}


\section{\label{sec:app_ge}Derivation of the Variational Equations}

Below we explicitly carry out the variation of the QCD vacuum energy density with respect
to the variational kernels $\omega$, $K_0$, $K$ of our trial ansatz [see Eqs.~\eqref{ans11}--\eqref{ans2}]
for the QCD vacuum wave functional.

\subsection{The Gluon Gap Equation}

If the implicit dependence of the ghost and quark loop, $\chi$ and $\sigma$, on the
gluon propagator is ignored,
from the gluon CRDSE~\eqref{gldse2} it is seen that the variation with respect to
$\omega$ can be traded for the variation with respect to $\Omega$ or, more conveniently,
with respect to the gluon propagator $D$ [\Eqref{G11}].

From the quark CRDSE~\eqref{qdse2} we have
\be\label{G20}
\frac{\delta Q^{-1}(1,2)}{\delta D(3,4)} = - \bar{\Gamma}_0(1,5;3) Q(5,6) \bar\Gamma(6,2;4)
\ee
resulting in
\be\label{G20a}
\frac{\delta E_\mathrm{D}^{(0)}}{\delta D(3,4)}
= h_0(2,1) Q(1,1') \frac{\delta Q^{-1}(1',2')}{\delta D(3,4)} Q(2',2)
= - h_0(2,1) Q(1,1') \bar{\Gamma}_0(1',5;3) Q(5,6) \bar\Gamma(6,2';4) Q(2',2) .
\ee
Minimization of $E_\mathrm{QCD}$ with respect to $D(1,2)$ then yields the condition
\begin{multline}\label{G4new}
\Bigl[-\frac12 D^{-1}(3,1) D^{-1}(2,4) \Bigr] D(4,5) \Bigl[ \frac12 D^{-1}(5,3)-\chi(5,3) \Bigr] \\
+\frac12 \Bigl[ \frac12 D^{-1}(3,1)-\chi(3,1) \Bigr] \Bigl[ \frac12 D^{-1}(2,3)-\chi(2,3) \Bigr]
-\frac12 \Delta(2,1)\\
- h_0(3,4) Q(4,1') \bar{\Gamma}_0(1',5;1) Q(5,6) \bar\Gamma(6,3';2) Q(3',3)
- J(3,4;1) Q(4,5) \bar\Gamma(5,6;2) Q(6,3) = 0
\end{multline}
Expressing the gluon propagator \Eqref{G11} by the gluon energy $\Omega$ this equation
can be simplified to
\[
\Omega^2(1,2) = -\Delta(1,2) + \chi^2(1,2) - 2 \bar\Gamma(5,6;1)Q(6,3)J(3,4;2) Q(4,5)
-2 \bar\Gamma(6,3';1) Q(3',3) h_0(3,4) Q(4,4') \bar\Gamma_0(4',5;2) Q(5,6) .
\]
If we suppress the fermionic indices we arrive at \Eqref{G4A}.


\subsection{The Quark Gap Equation}

The energy depends on the scalar kernel $K_0$ only through the quark propagator $Q$.
Hence the variation with respect to $K_0$ and $K_0^\dag$ can be carried out as
\be\label{G6}
\frac{\delta E_\mathrm{QCD}}{\delta K_0(2,1)} = \frac{\delta E_\mathrm{QCD}}{\delta Q(4,3)} \frac{\delta Q(4,3)}{\delta K_0(2,1)} \overset{!}{=} 0 ,
\qquad
\frac{\delta E_\mathrm{QCD}}{\delta K_0^\dag(2,1)} = \frac{\delta E_\mathrm{QCD}}{\delta Q(4,3)} \frac{\delta Q(4,3)}{\delta K_0^\dag(2,1)} \overset{!}{=} 0 .
\ee
From the quark CRDSE~\eqref{qdse2} and \Eqref{xG9} we have
\[
\frac{\delta Q(4,3)}{\delta K_0(2,1)} = - Q(4,4') \Lambda_+(4',2) \Lambda_-(1,3') Q(3',3) ,
\qquad
\frac{\delta Q(4,3)}{\delta K_0^\dag(2,1)} = - Q(4,4') \Lambda_-(4',2) \Lambda_+(1,3') Q(3',3) ,
\]
where we have included only the explicit $K_0$ dependence, since the implicit $K_0$ dependence
of $Q$ in the last term of \Eqref{qdse2} would lead to two-loop terms in \Eqref{G6}.
Defining
\be\label{G6A}
\frac{\delta E_\mathrm{QCD}}{\delta Q(2,1)} = -h(1,2)
\ee
the stationary condition \Eqref{G6} becomes
\be\label{G6B}
\Lambda_-(1,3) Q(3,3') h(3',4') Q(4',4) \Lambda_+(4,2) = 0 ,
\qquad
\Lambda_+(1,3) Q(3,3') h(3',4') Q(4',4) \Lambda_-(4,2) = 0 .
\ee
The quantity $h(1,2)$ [\Eqref{G6A}] defines an effective quasi-particle Hamiltonian
of the quarks. Restricting ourselves also up to including one-loop terms in the quark
gap equation \eqref{G6B} only those terms contribute to $h(1,2)$ which explicitly
depend on the quark propagator, i.e., the quark energy \Eqref{1447X21A} of \eqref{lastone}.
We find then
\begin{multline*}
h(1,2) = h_0(1,2) + J(1,3;4) Q(3,3') \bar\Gamma(3',2;4') D(4,4')
+ \bar\Gamma(1,3;4) Q(3,3') J(3',2;4') D(4,4') \\
+\frac18 \Bigl\{ \bar\Gamma_0(1,3;4) Q(3,3') \bar\Gamma(3',2;4) + \bar\Gamma(1,3;4) Q(3,3') \bar\Gamma_0(3',2;4)
- \bigl( \bar\Gamma_0 \to \bar\Gamma_-, \bar\Gamma \to \bar\Gamma_- \bigr) \Bigr\} \\
+ g^2 F(4,4') \bar{R}(1,3;4) \bigl[Q(3,3') - \tfrac12 Q_0(3,3') \bigr] \bar{R}(3',2;4') ,
\end{multline*}
where $\bar\Gamma_-$ is defined by \Eqref{nG19}. Using \Eqref{186x1} we recover \Eqref{G6Bxxx}.

Equations \eqref{G6B} are matrix-valued equations. Since $\Lambda_+ \Lambda_-=0$, the expressions on the
l.h.s.~of these equations are manifestly traceless.
The relevant information can be extracted by multiplying these equations with Dirac matrices and taking
the trace. All considerations given above are valid for massive bare quarks. The quark
gap equations~\eqref{G6B} simplify for massless bare quarks. For instance,
multiplying Eqs.~\eqref{G6B} with $\beta$ and taking the trace, thereby using $\beta \Lambda_+ = \Lambda_- \beta$,
which is valid for massless bare quarks,
we obtain the two conditions
\[
\Tr( Q h Q \Lambda_\pm \beta) = 0
\]
which can be collected in
\[
\Tr(Q h Q \beta ) = 0 .
\]


\subsection{\label{app:vkg}The Equation of Motion for the Vector Kernel}

Finally we derive the equation of motion for the vector kernel $K(1,2;3)$ [\Eqref{ans2}]
of the quark wave functional [Eqs.~\eqref{qcd0}, \eqref{dir5}]. The energy depends explicitly
on the vector kernel $K$ through the bare and full quark-gluon vertex
[$\bar\Gamma_0$ \Eqref{ans9} and $\bar\Gamma$ \Eqref{qgv} respectively],
and implicitly through the quark propagator $Q$. Restricting ourselves to one-loop
terms in the equation of motion we can neglect this implicit dependence in all energy
terms except in the free single-particle energy [first term on the r.h.s~of \Eqref{G16A}].
From the quark CRDSE~\eqref{qdse2} we get
\[
\frac{\delta Q^{-1}(6,7)}{\delta K(1,2;3)} =
- \frac{\delta \bar\Gamma_0(6,6';8)}{\delta K(1,2;3)} \, Q(6',7') \, D(8,8') \, \bar\Gamma(7',7;8')
- \bar\Gamma_0(6,6';8) \, Q(6',7') \, D(8,8') \, \frac{\delta \bar\Gamma(7',7;8')}{\delta K(1,2;3)} \, .
\]
The variation of the energy with respect to $K(1,2;3)$ yields therefore the following
equation of motion (for $K^\dag$)
\begin{align}\label{stst}
0 = {}&
h_0(4,5)Q(5,6)
\left[
\frac{\delta \bar\Gamma_0(6,6';8)}{\delta K(1,2;3)} \, Q(6',7') \, \bar\Gamma(7',7;8')
+ \bar\Gamma_0(6,6';8) \, Q(6',7') \, \frac{\delta \bar\Gamma(7',7;8')}{\delta K(1,2;3)}
\right] Q(7,4) \, D(8,8') \nn\\
&+ J(4,5;6) Q(5,7) \frac{\delta \bar\Gamma(7,8;9)}{\delta K(1,2;3)} Q(8,4) D(6,9)\nn \\
&+ \frac18 \left[ \frac{\delta \bar\Gamma_0(4,5;6)}{\delta K(1,2;3)} Q(5,7) \bar\Gamma(7,8;6) 
+ \bar\Gamma_0(4,5;6) Q(5,7) \frac{\delta \bar\Gamma(7,8;6)}{\delta K(1,2;3)} \right] Q(8,4) \nn\\
&+ \frac14 Q_0(4,5) \frac{\delta \bar\Gamma_0(5,6;7)}{\delta K(1,2;3)} Q(6,8)  \bar\Gamma_0(8,9;7) Q_0(9,10) Q(10,4)
\end{align}
For the bare quark-gluon vertex we find from its definition
\Eqref{ans9}
\be\label{stX}
\frac{\delta \bar\Gamma_0(4,5;6)}{\delta K(1,2;3)} = \Lambda_+(4,1) \delta(6,3) \Lambda_-(2,5),
\qquad
\frac{\delta \bar\Gamma_0(4,5;6)}{\delta K^\dag(1,2;3)} = \Lambda_-(4,1) \delta(6,3) \Lambda_+(2,5) .
\ee
On the other hand, from the CRDSE~\eqref{qgvdse2} for the full quark-gluon vertex $\bar\Gamma$,
which in the condensed notation of Sec.~\ref{sec:toten} reads
\be\label{st}
\bar\Gamma(1) = \bar\Gamma_0(1)
+ \bar\Gamma_0(2) Q \bar\Gamma(1) Q \bar\Gamma(3) D(2,3)
+ \bar\Gamma_0(2) Q \bar\Gamma(3) D(2,2') D(3,3') \Gamma(2',3',1)
- \bar\Gamma_0(2) Q \bar\Gamma_{\bar{q}qAA}(1,3) D(2,3)
\ee
we find the variation of the full quark-gluon vertex $\bar\Gamma$ with
respect to the vector kernels $K$ and $K^\dag$. In taking the variation of this equation
with respect to $K$, $K^\dag$ on the right-hand side we can replace the variation of the full vertices
by those of the bare ones ($\delta\bar\Gamma/\delta K \to \delta\bar\Gamma_0/\delta K$).
This is because the full vertices occur on the right-hand side of \Eqref{st} only inside
loops and the variation of their dressings would result in more than one loop. Since the variation
if the bare vertices are explicitly known [see~\Eqref{stX}], \Eqref{st} provides us with
an explicit expression for the variation of the full quark-gluon vertex, which has then
to be inserted into \Eqref{stst}. This completes the derivation of the variational equations
for $K$ and $^\dag$. For illustrative purposes we present here these equations also in
the bare-vertex approximation,
replacing the full vertex $\bar\Gamma$ by the bare one $\bar\Gamma_0$.
The equation of motion \eqref{stst} reduces then to
\begin{align}\label{ststst}\nn
0 = \Lambda_- Q \bigl\{ {}&D(1,2) \bigr[ \bar\Gamma_0(2) Q h_0 + h_0 Q \bar\Gamma_0(2) + J(2) \bigr] \\
&+ \frac18 \bigl[ 2 \bar\Gamma_0(1) + \bar\Gamma_0(1) Q_0 - Q_0\bar\Gamma_0(1) \bigr] \bigr\}Q \Lambda_+ .
\end{align}
Using \Eqref{186x1} we can rewrite the last term as
\[
2 \bar\Gamma_0(1) + \bar\Gamma_0(1) Q_0 - Q_0\bar\Gamma_0(1) = 2 (1-Q_0) \bar\Gamma_0(1) = 4\Lambda_- \bar\Gamma_0(1)
\]
where we have used the definition \eqref{bareprop} of $Q_0$ in terms of the projectors
$\Lambda_\pm$.


\section{\label{sec:app_en}The stationary energy}

For later application we simplify the expression for the energy at the stationary point.
For this purpose we multiply \Eqref{G20} with $D(3,4)$
and use the quark CRDSE~\eqref{qdse2} to find
\[
\frac{\delta Q^{-1}(1,2)}{\delta D(3,4)} D(3,4) = -\bar\Gamma_0(1,5;3) Q(5,6) \bar\Gamma(6,2;4) D(3,4)
= Q^{-1}(1,2) - Q_0^{-1}(1,2) - \bar\gamma(1,2) .
\]
With this relation we obtain from \Eqref{G20a}
\be\label{G20b}
\frac{\delta E_\mathrm{D}^{(0)}}{\delta D(3,4)} D(3,4) = h_0(2,1) \bigl\{ Q(1,2) - Q(1,1') \bigl[ Q_0^{-1}(1',2') + \bar\gamma(1',2') \bigr] Q(2',2) \bigr\} .
\ee
Multiplying now the gap equation \eqref{G4new} with $D(1,2)$ and using \Eqref{G20b}
we can express the sum of the Yang--Mills energy $E_\mathrm{YM}$ and quark-gluon
interaction energy $E_\mathrm{D}^{(1)}$ as
\[
\begin{split}
E_\mathrm{YM}+E_\mathrm{D}^{(1)} &\equiv E_E^\mathrm{YM} + E_B + E_\mathrm{D}^{(1)} \\
&= \frac12 \bigl[ \Omega(1,1)-\chi(1,1) \bigr]- h_0(1,2) Q(2,1) + h_0(2,2') Q(2',1') \bigl[ Q_0^{-1}(1',1) + \bar\gamma(1',1) \bigr] Q(1,2) ,
\end{split}
\]
where we have used in the last expression the quark CRDSE~\eqref{qdse2}. Adding here also the
free Dirac energy $E_\mathrm{D}^{(0)}$ we obtain
\be\label{G20c}
E_\mathrm{YM} + E_\mathrm{D} = \frac12 \bigl[ \Omega(1,1) - \chi(1,1) \bigr]
-h_0(2,1) \bigl[ 2Q(1,2) - \Lambda_+(1,2) \bigr]
+ h_0(1,1') Q(1',2') \bigl[ Q_0^{-1}(2',2) + \bar\gamma(2',2) \bigr] Q(2',2) \bigr] .
\ee
Note that this expression holds only at the stationary point [i.e.,~for gluon propagators
satisfying the gap equation \eqref{G4A}].
Formally, the first term on the r.h.s.~of \Eqref{G20c} is the same as the one obtained in Ref.~\cite{Heffner:2012sx} for
the pure Yang--Mills sector. However, in the present case $\Omega$ is the solution
of the gap equation \eqref{G4A}, which contains the quark loop. Also $\chi$ [\Eqref{ghostloop}],
being a functional of $\Omega$ through the ghost propagator, will, of course, be different.


\bibliographystyle{h-physrev5}{}
\bibliography{biblio-spires}

\end{document}